\documentclass[twocolumn,showpacs,preprintnumbers,amsmath,amssymb]{revtex4}

\usepackage{graphicx}
\usepackage{dcolumn}
\usepackage{bm}
\usepackage{amssymb}
\usepackage{latexsym}
\usepackage{amsmath}
\usepackage{epsfig}
\usepackage{tabularx}
\usepackage{amsmath}
\usepackage{pict2e,booktabs,amsmath}
\usepackage[usenames,dvipsnames,svgnames]{xcolor}

\newcommand{\ket}[1]{\ensuremath{|\,#1\,\rangle}}

\begin{document}


\title{Model for the hyperfine structure of electronically-excited ${\rm KCs}$ molecules }
\author{A.~Orb\'an $^{1}$}
\author{R.~Vexiau$^{1}$}
\author{O.~Krieglsteiner $^{2}$}
\author{H.-C. N\"agerl$^{2}$}
\author{O.~Dulieu$^{1}$}
\author{A.~Crubellier$^{1}$}
\author{N.~Bouloufa-Maafa$^{1}$}

\affiliation{$^{1}$Laboratoire Aim\'e Cotton, CNRS/ Universit\'e Paris-Sud/ ENS Cachan, B\^at. 505, Campus d'Orsay, 91405 Orsay Cedex, France}
\affiliation{$^{2}$Institut f\"ur Experimentalphysik und Zentrum f\"ur Quantenphysik Universit\"at Innsbruck\\
Technikerstra\ss e 25, 6020 Innsbruck, Austria }

\date{\today}

\begin{abstract}
A model for determining the hyperfine structure of the excited electronic states of diatomic bialkali heteronuclear molecules is formulated from the atomic hyperfine interactions, and is applied to the case of bosonic $^{39}$KCs and fermionic $^{40}$KCs molecules. The hyperfine structure of the potential energy curves of the states correlated to the K($4s\,^2S_{1/2}$)+Cs($6p\,^2P_{1/2,3/2}$) dissociation limits is described in terms of different coupling schemes depending on the internuclear distance $R$. These results provide the first step in the calculation of the hyperfine structure of rovibrational levels of these excited molecular states in the perspective of the identification of efficient paths for creating ultracold ground-state KCs molecules.
\end{abstract}
\maketitle

\section{Introduction}
\label{sec:intro}

The control of the evolution of atomic and molecular systems at the single quantum level is an opportunity that is offered to fundamental physics by the amazing developments in the research on ultracold quantum gases, \textit{i.e.} where particles are moving with kinetic energies $E=k_BT$ equivalent to temperatures $T$ much smaller than 1 millikelvin. Among the most spectacular achievements are the observation of quantum degeneracy in ultracold atomic gases of bosons \cite{anderson1995,davis1995,bradley1995} and fermions \cite{demarco1999}. Quantum degeneracy and even Bose-Einstein condensation of weakly-bound ultracold molecules created by the association of pairs of atomic bosons or fermions has also been achieved \cite{heinzen2000,donley2002,jochim2003a,herbig2003,zwierlein2003,greiner2003}, with the major drawback that dimer molecules are quite unstable against collisions as they contain a lot of internal energy, unless they are protected by the Pauli principle as in the case of fermion pairs. Major progress towards the formation of a quantum gases of ultracold molecules in their absolute ground state has been reported since 2008 \cite{ni2008,danzl2008,lang2008a,danzl2010,takekoshi2014,molony2014}, and the observation of quantum degeneracy is now within reach. In most cases this research relies on two main steps: (i) the formation of weakly-bound molecules by association of a pair of ultracold atoms via a Feshbach resonance \cite{kohler2006,chin2010}, and (ii) the transfer of the population from this so-called Feshbach state toward neighboring levels using avoided crossings \cite{mark2007b} or radiofrequency transitions \cite{lang2008}, or toward the absolute ground level of the molecule using a coherent optical process known as stimulated Raman adiabatic passage (STIRAP) \cite{bergmann1998,vitanov2001,koch2012}. Alternative methods to obtain molecular degenerate gases are also under investigation starting from preexisting molecules, like evaporative cooling \cite{stuhl2012}, buffer gas cooling \cite{egorov2004,hutzler2012}, Sisyphus cooling \cite{zeppenfeld2012} or laser cooling \cite{dirosa2004,hummon2013,zhelyazkova2014,barry2014,kobayashi2014}.

All the above-mentioned methods require a precise control of the internal state of the molecule of interest. In particular the STIRAP technique has proven to be  very versatile for controlling the internal state in all degrees of freedom (vibrational, rotational, and hyperfine) of ultracold diatomic molecules composed of identical atoms \cite{danzl2010,lang2008a} or of different alkali atoms \cite{ni2008,takekoshi2014,molony2014}. This has allowed the creation of \textit{polar} molecules, which are currently of great interest \cite{doyle2004,carr2009,dulieu2009,jin2012}. In brief, STIRAP is based on a so-called \textit{Lambda} scheme of energy levels: the population is transferred from an initial level \ket{i} toward a final level \ket{g} via two overlapping laser pulses involving an intermediate level \ket{e} that is never populated. For ultracold polar bialkali molecules, the scheme is implemented as follows: the \ket{i} level is a weakly-bound level of the molecular ground state manifold (usually referred to as a Feshbach molecule) with mixed $^3\Sigma^+$ and $^1\Sigma^+$ symmetries, the \ket{g} is the absolute ground level of the molecular ground state X$^1\Sigma^+$, and the \ket{e} levels belongs to an electronically excited state chosen such that it has noticeable dipole-allowed transition probabilities with both \ket{i} and \ket{g}. 

The knowledge of the tiniest properties of the quantum states involved in a STIRAP scheme - namely the hyperfine structure (hfs)- is essential to ensure the optimal efficiency of the population transfer. The \ket{i} hfs is usually well-known for the alkali-atom pairs, at it is mostly determined by the hfs of the separated ground state $^2S$ atoms involving only their electronic spin quantum number $s=1/2$ and their nuclear spin quantum number $i_n$. It has been largely used for the modeling of Fano-Feshbach resonances (FFR) in mixed alkali-metal atom pairs (\textit{i.e.} LiNa \cite{stan2004}, LiK \cite{wille2008}, LiRb \cite{deh2008,marzok2009}, LiCs \cite{repp2013,cho2013}, NaK \cite{wu2012}, NaRb \cite{wang2013}, KRb \cite{inouye2004}, KCs \cite{patel2014}, RbCs \cite{pilch2009}). On the other hand very little is known on the hfs of ground-state heteronuclear alkali dimers in their lowest rovibrational levels (the \ket{g} state above), with a few remarkable exceptions  \cite{aldegunde2008,aldegunde2009,ran2010,ospelkaus2010b,strauss2010,wall2013}.

The hfs of the \ket{e} levels is intrinsically more complicated as it involves one $^2P$ atom, \textit{i.e.} one more non-zero angular momentum to be coupled. Its spectroscopic observation could also be hindered by the natural width of the levels. Several studies have extended the long-range analysis of the hfs of a pair of $^2S$ ground-state alkali-metal atoms to the particular case where pure long-range potentials wells \cite{stwalley1978,movre1977} are present in the electronic states dissociating to the lowest $^2S+^2P$ asymptote of homonuclear alkali dimers \cite{comparat2000,kemmann2004,tiesinga2005}. Molecular spectroscopy has allowed the investigation of deeply-bound vibrational levels of excited states of NaRb \cite{kasahara1996} and NaK \cite{burns2003,burns2005,wilkins2005}. In the ultracold regime high-resolution spectroscopy has allowed observing and studying the hfs for low-lying vibrational levels of one of the first excited states of Rb$_2$ \cite{takekoshi2011} and RbCs \cite{debatin2011}, which have been modeled with a simple effective Hamiltonian involving parameters fitted to the measurements. Last but not least, \textit{ab initio} molecular hyperfine parameters have been computed for the lowest $^3\Sigma^+$ states of homonuclear alkali dimers \cite{lysebo2013}.

Despite all the achievements above, the complexity of the problem may have prevented up to now the complete description of the hfs of excited molecular states, as a prerequisite is the determination of complete potential energy curves (PECs) including hfs. In this paper, we extend the asymptotic model developed in Ref.~\cite{comparat2000} to all molecular states correlated to the two lowest $^2S+^2P$ asymptotes of polar alkali-metal diatomic molecules. We illustrate it by deriving PECs including fine and hyperfine structure for all internuclear distances $R$ for the states of $^{39}$KCs and $^{40}$KCs correlated to K($4s\,^2S_{1/2}$)+Cs($6p\,^2P_{j}$) (with $j=1/2,3/2$). We first recall the expression of the full Hamiltonian for the atom pair including spin-orbit and hyperfine couplings, and we detail the data used for the calculation for both the bosonic and the fermionic isotopologues. The variation of coupling regime along the PECs when $R$ varies is discussed in terms of "good" quantum numbers, shedding some light on the expected level structure depending on its binding energy. 

\section{The Hamiltonian for a pair of K and C{s} atoms}
\label{sec:hamiltonian}

\begin{figure}
\includegraphics[scale=0.5]{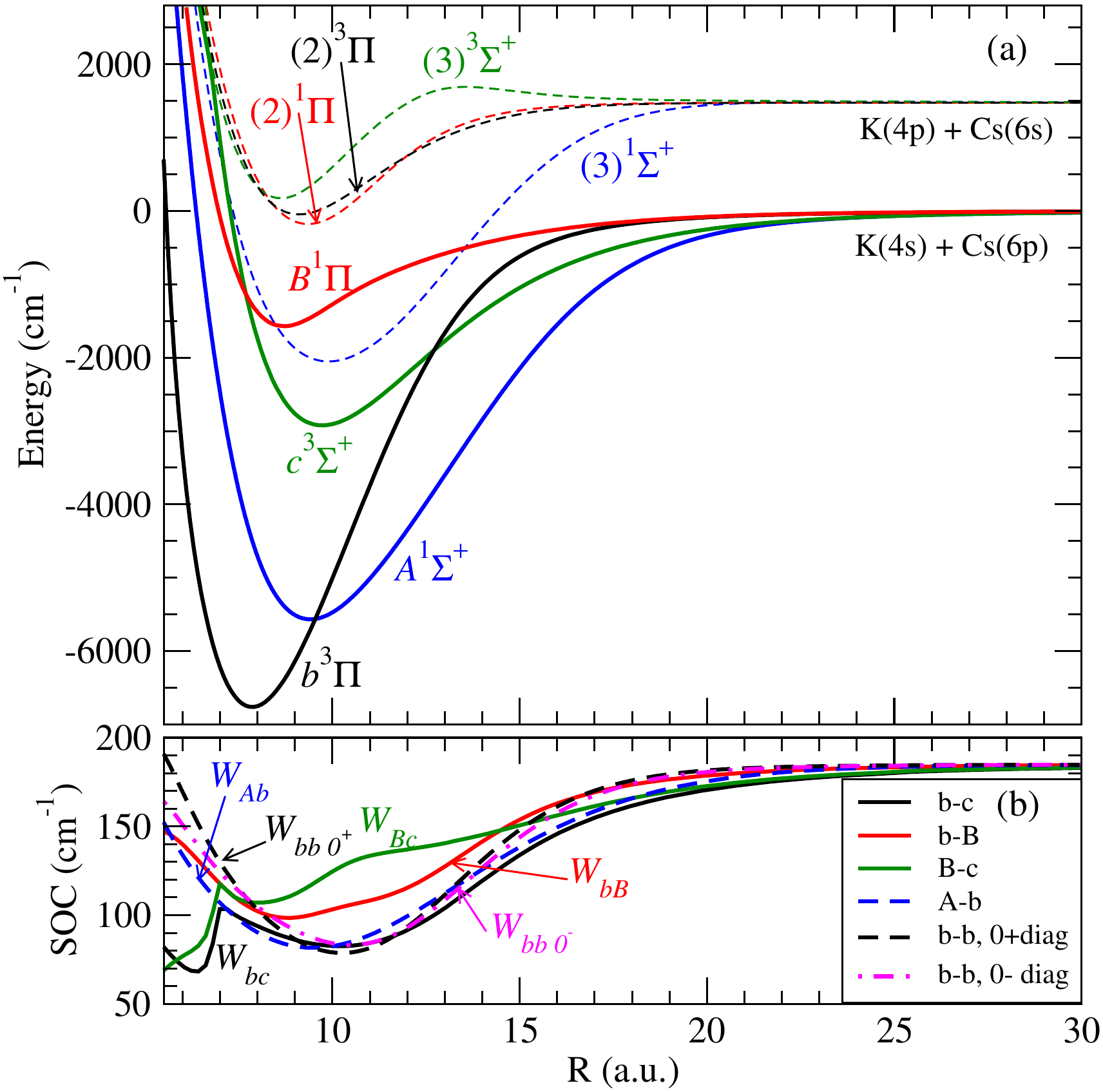}
\caption{(a) Hund's case {\it a} potential energy curves of the excited electronic states correlating to the K$(4^2S)$+Cs$(6^2P)$ (solid lines) and K$(4^2P)$+Cs$(6^2S)$ (dashed lines) dissociation limits. The $B$ and $c$ curves cross at $R^{Bc}=7.65$~a.u., the $b$ and $A$ curves at $R^{bA}=9.55$~a.u., and the $b$ and $c$ states at $R^{bc}=12.84$~a.u.; (b) the relevant $R$-dependent spin-orbit coupling terms associated to the curves dissociating to K$(4^2S)$+Cs$(6^2P)$ (1~a.u.~$\equiv a_0=0.0529177$~nm).} 
\label{fig:hundsa}
\end{figure}

Our goal is to compute Born-Oppenheimer (BO) PECs including fine and hyperfine structure. Thus the rotation of the molecule is not considered at that step. Within the standard BO approximation we write the electronic Hamiltonian of the system as
\begin{equation}
\hat{{\bf H}}=\hat{{\bf H}}^0+\hat{{\bf H}}^{\text{f}}+\hat{{\bf H}}^{\text{hf}}
\label{eq:total}
\end{equation}
where the electrostatic Hamiltonian $\hat{\bf H}^0$ containing the kinetic energy of the electrons and the sum of the potential energies of charged particles is perturbed by the fine structure $\hat{\bf H}^{\text{f}}$ and the hyperfine structure $\hat{\bf H}^{\text{hf}}$ Hamilton operators. We assume the electrostatic problem solved so that the matrix of $\hat{\bf H}^0$ is diagonal and contains the $R$-dependent PECs in Hund's case \textit{a} (see Fig.~\ref{fig:hundsa} for the KCs curves correlated to K($4s\,^2S$)+Cs($6p\,^2P$)).

For $\hat{\bf H}^{\text{f}}$ we only consider the molecular spin-orbit (SO) interaction, which is formally written as:
\begin{equation}
\hat{{\bf H}}^{\text{f}}=A(R){\bf L} . {\bf S}
\label{eq:HSO}
\end{equation}
where ${\bf L}$ and ${\bf S}$ are the total electronic orbital momentum and spin. The $R$-varying coupling constant $A(R)$ depends on the considered molecular states and is taken either from available spectroscopic data or from quantum chemistry calculations, as discussed below.

Following Ref.~\cite{comparat2000}, we apply here the main approximation of our model: we express $\hat{\bf H}^{\text{hf}}$ as the sum of two atomic hyperfine Hamiltonian operators $\hat{{\bf h}}^{\text{hf}}(k)$ ($k=1,2$), which characterize the experimental hyperfine structure of the individual atoms. In other words, we assume that the related coupling constants do not vary with $R$, and we define them in order to reproduce the experimental hf splitting of the atoms. In the absence of published computations of the appropriate molecular constants --which are outside the scope of the present paper-- this approximation is reasonable. For instance the variation of these coupling constants is expected to be limited to about $\approx$15\% in the case of RbSr ground state \cite{zuchowski2010a}. A similar trend is also found for the hfs of the $1^3\Sigma_u^+$ and $1^3\Sigma_g^+$ states of alkali-metal dimers \cite{lysebo2013}. Such an approximate Hamiltonian thus mimics the interaction between the electrons and the nucleus within each atom usually expressed in terms of multipolar interactions, with odd magnetic and even electronic multipoles due to parity conservation inside the nucleus \cite{broyer1978}:
\begin{equation}
\hat{{\bf h}}^{\text{hf}}=\hat{\bf h}^{\text{hf}}_{MD}+\hat{\bf h}^{\text{hf}}_{EQ}=\hat{\bf h}^{\text{hf}}_{{\bf \ell i}}+ \hat{\bf h}^{\text{hf}}_{{\bf s i}}+\hat{\bf h}^{\text{hf}}_{{\bf FC}}+\hat{\bf h}^{\text{hf}}_{EQ}
\label{eq:HFS}
\end{equation}
with the symbols $\ell$, $s$, and $i$ referring to the atomic angular momenta. The two main terms are: (i) the magnetic dipole interaction Hamiltonian $\hat{\bf h}^{\text{hf}}_{MD}$, which incorporates the interaction of the nuclear spin with the orbital angular momenta of the electrons $\hat{\bf h}^{\text{hf}}_{{\bf \ell i}}$ and with the spin of the electrons $\hat{\bf h}^{\text{hf}}_{{\bf s i}}$, completed by the Fermi contact term $\hat{\bf h}^{\text{hf}}_{{\bf FC}}$ induced by the electrons passing through the nuclei; (ii) the electric quadrupole term $\hat{\bf h}^{\text{hf}}_{EQ}$, which describes the interaction between the electric quadrupole moment of the nucleus with the gradient of the electric field created by the electrons at the nucleus. In our model the interaction of the electronic and nuclear momenta of one atom with those of the other atom is neglected, so that we do not consider the corresponding multipolar terms.

We apply the model to the first excited states of $^{39}{\rm K}^{133}{\rm Cs}$ and $^{40}{\rm K}^{133}{\rm Cs}$ molecules, which tend  to the K($4^2S$)+Cs($6^2P$) dissociation limit. Four electronic states are defined in the Hund's case {\it a} notation: $A^1\Sigma^+$, $c^3\Sigma^+$, $B^1\Pi$ and $b^3\Pi$. Their potential energy curves (PECs) are presented on Fig.~\ref{fig:hundsa}. They come either from the spectroscopic study of Refs.~\cite{kruzins2010,kruzins2013}, or from our own quantum chemical calculations based on effective core potentials (ECPs) and core polarization potentials (CPPs) implemented in a full configuration interaction approach (FCI) built on a large Gaussian basis set \cite{aymar2005,aymar2006a,guerout2010}. More details about the construction of the PECs can be found in \cite{borsalino2015}. Note that three out of four PECs dissociating to the next asymptote, K($4^2P$)+Cs($6^2S$) do not cross the PECs connected to the K($4^2S$)+Cs($6^2P$) lower limit. The (3)$^1\Sigma^+$ and the $B^1\Pi$ PECs cross each other but these states have no direct spin-orbit interaction, and we will not include the former state in the present investigation.

The chosen model for the interaction Hamiltonians above suggests to employ the well-known technique of linear combination of atomic orbitals to derive the expressions of the coupling matrix elements of the spin-orbit and hyperfine interactions. We thus consider the basis $\left|L, S, \Lambda, m_S \right\rangle$ where $L$ and $S$ are the total electronic angular momentum and spin quantum numbers, and $\Lambda$ and $m_S$ the quantum numbers for their projection on the molecular axis:
\begin{eqnarray}
\label{eq:atbas}
\left|L, S, \Lambda, m_S \right\rangle&=&\frac{1}{2}\left( \,\,\, 
                   \left| l_1m_1\right.\rangle_{\rm K}^1\left| l_2m_2\right.\rangle_{\rm Cs}^2\right. \nonumber\\
     &&\left.+(-1)^{S}\left| l_1m_1\right.\rangle_{\rm K}^2\left| l_2m_2\right.\rangle_{\rm Cs}^1 \,\,\,\right )\left|S \,m_S \right.\rangle^{1,2}.\nonumber\\
     &&
\end{eqnarray}
The ket $\left|S\,m_S \right.\rangle^{1,2}$ is the coupled spin-function of the two valence electrons, and $ \left| l_1m_1\right.\rangle$ and $ \left| l_2m_2\right.\rangle$ stand for the $s$ and $p$ atomic orbitals with $l_1=0$, $m_1=0$ and $l_2=1$, $m_2=-1,0,1$. For the present asymptotic limit we have $L=1$ and $\Lambda=m_2$, resulting in 12 basis functions $\left|L, S, \Lambda, m_S \right\rangle$.

\subsection{Spin-orbit coupling}
\label{ssec:soc}
In the $\left|L, S, \Lambda, m_S \right\rangle$ basis the $\hat{{\bf H}}^{\text{f}}$ operator couples states with the same $ \Lambda+m_S=\Omega$ value, where $\Omega$ is the projection on the molecular axis of the total electronic angular momenta ${\bf j_e = L+S}$. As in the next step the hf interaction will couple most of the 12 states above, it is worthwhile to display the structure of $\hat{\bf H}^{\text{f}}$ operator in a compact and global way, as given in Eq.~(\ref{eq:sotable}):
\newline
\begin{equation}
\resizebox{0.49\textwidth}{!}{$
\label{eq:sotable}
\left(
\begin{array}{cc|cc|cccccc|cc}
\hspace*{1.2cm}{\bf 0^-}&&\hspace*{1.2cm}{\bf 0^+}&&&&\hspace*{1.2cm}{\bf 1}&&&&\hspace*{0.7cm}{\bf 2}&\\
{\bf ^3\Pi}^1_{-1}&{\bf ^3\Sigma}^0_{0}&
{\bf ^3\Pi}^1_{-1}&{\bf ^1\Sigma}^0_{0}&
{\bf ^3\Pi}^0_{1}&^3\Pi^0_{-1}&{\bf ^1\Pi}^0_{1}&^1\Pi^0_{-1}&{\bf ^3\Sigma}^1_{0}&^3\Sigma^{-1}_{0}&
{\bf ^3\Pi}^1_{1}&^3\Pi^{-1}_{-1}\\
-1&\sqrt{2}&0&0&\,\,0&\,\,\,0&\,\,\,0&\,\,\,0&\,\,\,0&\,\,\,0&\,\,\,0&\,\,\,0\\
\sqrt{2}&0&0&0&\,\,0&\,\,\,0&\,\,\,0&\,\,\,0&\,\,\,0&\,\,\,0&\,\,\,0&\,\,\,0\\
\hline
0&0&-1&\sqrt{2}&\,\,0&\,\,\,0&\,\,\,0&\,\,\,0&\,\,\,0&\,\,\,0&\,\,\,0&\,\,\,0\\
0&0&\sqrt{2}&0&0&\,\,\,0&\,\,\,0&\,\,\,0&\,\,\,0&\,\,\,0&\,\,\,0&\,\,\,0\\
\hline
0&0&\,\,0&0&\,\,0&\,\,\,0&\,\,\,1&\,\,\,0&\,\,\,1&\,\,\,0&\,\,\,0&\,\,\,0\\
0&0&\,\,0&0&\,\,0&\,\,\,0&\,\,\,0&-1&\,\,\,0&\,\,\,1&\,\,\,0&\,\,\,0\\
0&0&\,\,0&0&\,\,1&\,\,\,0&\,\,\,0&\,\,\,0&-1&\,\,\,0&\,\,\,0&\,\,\,0\\
0&0&\,\,0&0&\,\,0&-1&\,\,\,0&\,\,\,0&\,\,\,0&\,\,\,1&\,\,\,0&\,\,\,0\\
0&0&\,\,0&0&\,\,1&\,\,\,0&-1&\,\,\,0&\,\,\,0&\,\,\,0&\,\,\,0&\,\,\,0\\
0&0&\,\,0&0&\,\,0&\,\,\,1&\,\,\,0&\,\,\,1&\,\,\,0&\,\,\,0&\,\,\,0&\,\,\,0\\
\hline
0&0&\,\,0&0&\,\,0&\,\,\,0&\,\,\,0&\,\,\,0&\,\,\,0&\,\,\,0&\,\,\,1&\,\,\,0\\
0&0&\,\,0&0&\,\,0&\,\,\,0&\,\,\,0&\,\,\,0&\,\,\,0&\,\,\,0&\,\,\,0&\,\,\,1
\end{array}
\right)
$}
\end{equation}
\newline
For simplicity, this expression exhibits the location of the non-vanishing asymptotic ($R \rightarrow \infty$) spin-orbit matrix elements, which have to be multiplied by the Cs($6p$) spin-orbit splitting ($\Delta E_{\rm{so}}/3=184.679$~cm$^{-1}$) \cite{nist_database}. The diagonalization of this matrix yields the energies of atomic fine-structure levels. Each block is characterized by a value of $\left|\Omega \right|$ (first line header in Eq.~(\ref{eq:sotable}), with the index $\pm$ accounting for the reflection symmetry of $\Omega=0$ states) and involves Hund's case \textit{a} states (second line header), which are labeled as $^{2S+1}\Sigma_{\Lambda=0}^{m_S}$ and $^{2S+1}\Pi_{\Lambda=\pm1}^{m_S}$, with $m_S=0,\pm1$. States with positive values of $\Omega$ are written in boldface. The degeneracy of the states with $\pm \Omega \neq 0$ is not removed by the spin-orbit interaction.

\begin{figure}[bt]
\includegraphics[scale=0.45]{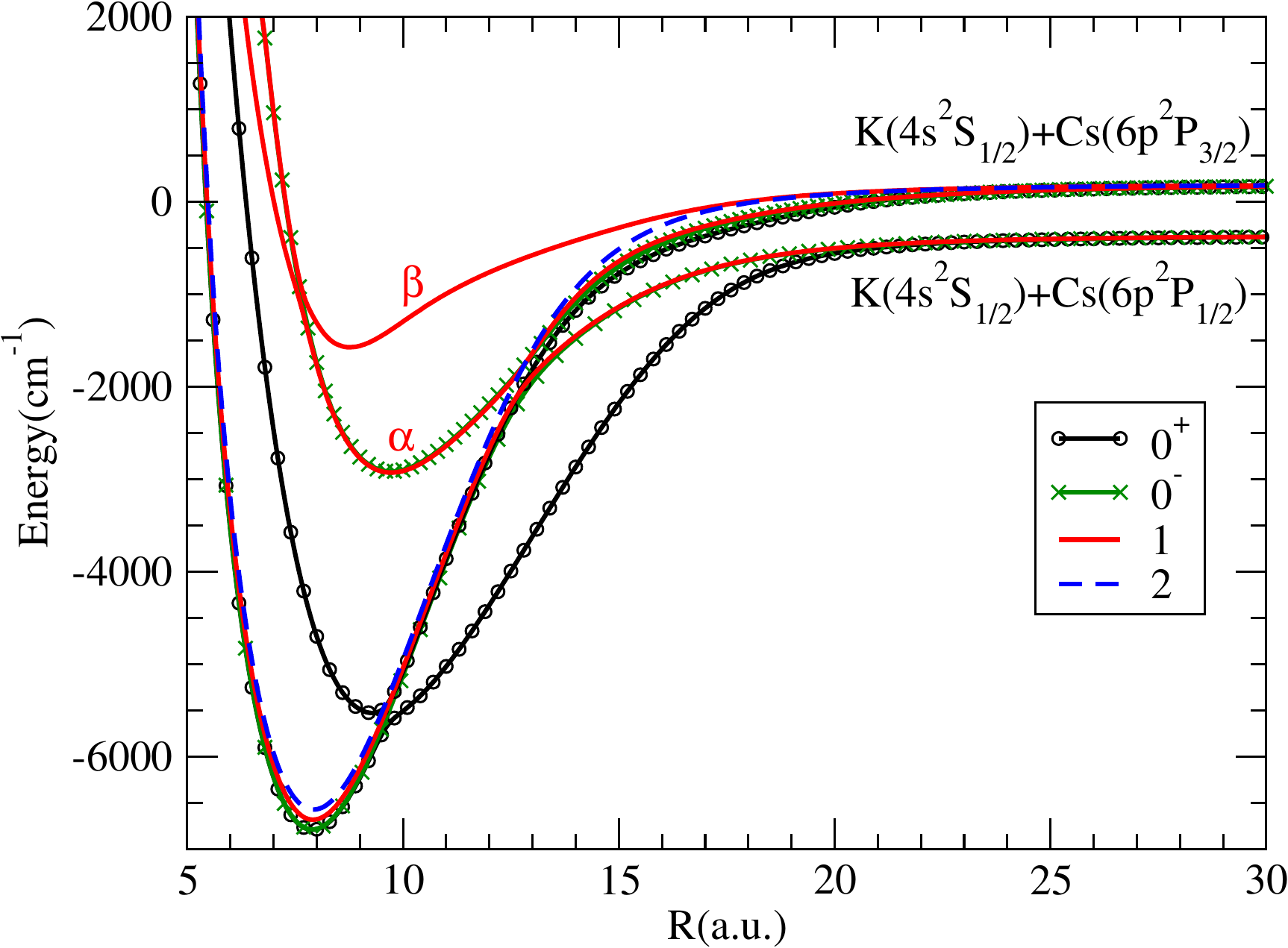}
\caption{Hund's case {\it c} potential energy curves of excited molecular states correlating to the dissociation limits K$(4^2S_{1/2})$+Cs$(6^2P_{1/2,3/2})$. The $\alpha$ and $\beta$ indexes label the two $\Omega=1$ states correlated to the upper dissociation limit.} 
\label{fig:hundsc}
\end{figure}

In order to obtain Hund's case \textit{c} PECs including spin-orbit, we have replaced the $R$-dependent spin-orbit matrix elements displayed in Fig.~\ref{fig:hundsa}(b) to the atomic constant in Eq.~(\ref{eq:sotable}). Note that all these functions indeed converge to $\Delta E_{\rm{so}}/3=184.679$~cm$^{-1}$ at large distances. They are taken from various experimental and theoretical papers \cite{tamanis2010,kruzins2010,kim2009}, as explained in detail in Ref.~\cite{borsalino2015}. The PECs resulting from the diagonalization of the Hamiltonian matrix including this molecular spin-orbit couplings are plotted in Fig.~\ref{fig:hundsc}. To identify states with the same $\Omega$ value we label them with a short notation expressing their dissociation limit: $2(P_{3/2})$, $1(P_{1/2})$, $1(P_{3/2}{\alpha})$, $1(P_{3/2}{\beta})$, $0^{+/-}(P_{1/2})$ and $0^{+/-}(P_{3/2})$ notations. The two $(P_{3/2})$ states with $\Omega=1$ are distinguished by the index $\alpha$ and $\beta$ for the lowest and the upper one, respectively.

\subsection{Hyperfine interaction}
\label{ssec:hyperfine}

We build the hyperfine Hamiltonian upon the hyperfine structure of the K and Cs atoms, which is recalled in Fig.~\ref{fig:atomichfs} and in Table~\ref{tab:HFS-atoms} in the Appendix. The hyperfine splitting of the $^{39,40}$K($4\,^2S_{1/2}$) and the $^{133}$Cs($6\,^2P_{1/2}$) state is determined only by the magnetic dipole term, the electric quadrupole and magnetic octupole terms being zero for $j=1/2$. The latter terms give non-zero contribution only for the $^{133}$Cs($6\,^2P_{3/2}$) state. In the following we will omit the mass index of the Cs atom, for simplicity. The nuclear spins are 3/2, 4 and 7/2 for $^{39}$K, $^{40}$K, and Cs, respectively. We set up the molecular basis by assembling the atomic states $\left| ((s_{k}l_{k})j_{k},i_{k})f_{k},m_{f_{k}}\right.\rangle $, where $i_{k}$ and $f_{k}$ denote the nuclear spin and the total angular momentum of the atom $k$, respectively. We use the simplified notation
\begin{eqnarray}
\label{eq:molbas}
 \left| (f_{1}, f_{2})f\, m_f\right.\rangle &\equiv& 
             \left| (((s_{1}l_{1})j_{1}i_{1})f_{1}, ((s_{2}l_{2})j_{2}i_{2})f_{2})f\, m_f\right.\rangle \nonumber\\
           &=&\sum_{m_{f_1},m_{f_2}} (-1)^{-f_1+f_2-m_{f}} \sqrt{2f+1} \nonumber\\
      &\times& \left(
                     \begin{array}{ccc}
                        f_1&f_2&f\\
                        m_{f_{1}}&m_{f_{2}}&-m_{f}
                     \end{array}
                \right) \nonumber \\
      &\times& \left| ((s_{1}l_{1})j_{1},i_{1})f_{1},m_{f_{1}}\right.\rangle \nonumber \\
      &\times& \left| ((s_{2}l_{2})j_{2},i_{2})f_{2},m_{f_{2}}\right.\rangle,
\end{eqnarray}
where $(...)$ denotes $3j$-symbols. The quantum numbers $f$ and $m_f$ are associated to the total angular momentum (without the molecular rotation) of the molecule ${\bf f}={\bf f_1}+{\bf f_2}$ and its projection on the molecular axis. At infinite separation we obviously have
\begin{eqnarray}
\label{hfsenergy}
\langle\left. (f_{1},f_{2})f\, m_f\right|H^{\text{hf}}\left| (f_{1},f_{2})f\, m_f\right.\rangle 
                                          &=&E^{\text{hf}}_1(i_1,j_1,f_1) \nonumber\\
                                          &+&E^{\text{hf}}_2(i_2,j_2,f_2),
\end{eqnarray}
$E^{\text{hf}}_1(i_k,j_k,f_k)$ denoting the energy of the $(i_k,j_k,f_k)$ hyperfine level of the atom $k$.

\begin{figure}[tb]
\includegraphics[scale=0.5]{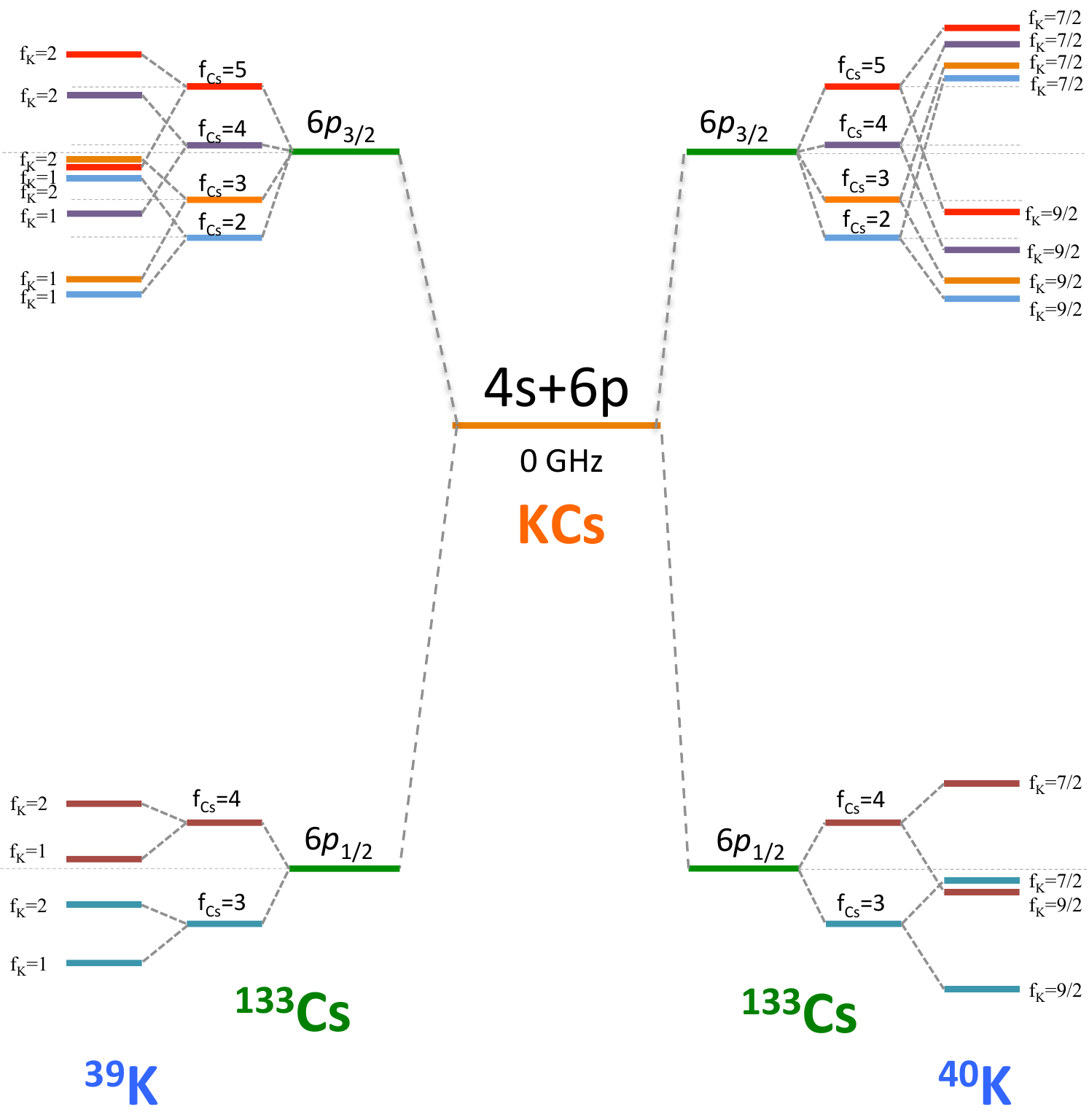}
\caption{Scheme of the asymptotic molecular hyperfine structure of the bosonic $^{39}$KCs and fermionic $^{40}$KCs molecules at the K($4\,^2S_{1/2}$)+Cs($6\,^2P_{1/2,3/2}$) dissociation limit. Energy spacings are not on scale, but the energy level order is meaningful. Exact energy splittings are reported in the Appendix.} 
\label{fig:atomichfs}
\end{figure}

The matrices of $\hat{\bf H}^{\text{0}}$ and $\hat{\bf H}^{{\text f}}$ in this basis are obtained after a basis transformation between the asymptotic molecular basis above and the $\left|L, S, \Lambda, m_S \right\rangle$ basis completed by the quantum numbers $\left|I,m_I \right\rangle$ for the total nuclear spin of the molecule and its projection $m_f=\Lambda +m_S +m_I$ on the molecular axis 
\begin{eqnarray}
\left| (f_{1},f_{2})f\, m_f\right.\rangle &=&
\sum_{\substack{ j,I,m_j,m_{I},\\S,L,m_S,m_L,\\m_{l_1},m_{l_2}}}
\hat{S}\hat{L}\hat{j_1}\hat{j_2}\hat{j}\hat{I}\hat{f_1}\hat{f_2}\nonumber \\
&&C_{j, m_j, I, m_I}^{f,m_f}C_{S, m_S, L, m_L}^{j,m_j} C_{l_1, m_{l_1}, l_2, m_{l_2}}^{L,m_L}\nonumber \\
&&
\left\{
 \begin{array}{ccc}
 j_1&i_1&f_1\\
 j_2&i_2&f_2\\
 j&I&f
 \end{array} 
\right\}
\left\{
 \begin{array}{ccc}
 s_1&l_1&j_1\\
 s_2&l_2&j_2\\
 S&L&j
 \end{array} 
\right\} \nonumber \\
&& \left|L, S, \Lambda, m_S \right\rangle \left|I m_I\right.\rangle,
\end{eqnarray}
where $\hat{X}\equiv \sqrt{2X+1}$, the factors $C$ are Clebsh-Gordan coefficients, and $\left\{...\right\}$ are $9j$ coefficients. We thus have $I=2,...,5$ for $^{39}$KCs and $I=\frac{1}{2},...,\frac{15}{2}$ for $^{40}$KCs.

We have defined in Section~\ref{ssec:soc} 12 $\left|L, S, \Lambda, m_S \right\rangle$ states, while 32 (resp. 72) nuclear spin states $\left|Im_I\right\rangle$ exist for $^{39}$KCs (resp. $^{40}$KCs), resulting in a total of 384 (resp. 864) $\left| (f_{1}, f_{2})f\, m_f\right.\rangle$ hyperfine states. As states with opposite values of $m_f$ are degenerate in energy in the absence of external field, it is sufficient to count states with $m_f \ge 0$ (Tables~\ref{tab:t39KCs} and ~\ref{tab:t40KCs}). For instance, the number of $^{39}$KCs states with $m_f=0$ is 48, so that there are (384-48)/2+48=216 different hyperfine states labeled with $m_f\ge0$. As the hf interaction is implemented as a perturbation of Hund's case c states, Tables~\ref{tab:t39KCs} and ~\ref{tab:t40KCs} also display the number of non-degenerate states for each $\Omega \ge 0$ symmetry. These numbers will be of great help to interpret the molecular PECs including hf interaction obtained in the next Section.

\begin{table}[htbp]
\scalebox{0.8}{
\begin{tabular}{c| l| lll| llll| l}
\hline
\toprule
\textbf{$\Omega$} &\makebox[3em]{2}&\makebox[3em]{1}&\makebox[3em]{1$^{\alpha}$}&\makebox[3em]{1$^{\beta}$}&\makebox[3em]{0$^+_{1/2}$}&\makebox[3em]{0$^-_{1/2}$}&\makebox[3em]{0$^+_{3/2}$}&\makebox[3em]{0$^-_{3/2}$}&{\bf Tot.}\\

\hline
\midrule
\textbf{$m_f$} &&&&&&&&\\
0&\makebox[3em]{8}&\makebox[3em]{8}&\makebox[3em]{8}&\makebox[3em]{8}&\makebox[3em]{4}&\makebox[3em]{4}&\makebox[3em]{4}&\makebox[3em]{4}&\,\,{\bf48}\\
1&\makebox[3em]{7}&\makebox[3em]{8}&\makebox[3em]{8}&\makebox[3em]{8}&\makebox[3em]{4}&\makebox[3em]{4}&\makebox[3em]{4}&\makebox[3em]{4}&\,\,{\bf47}\\
2&\makebox[3em]{6}&\makebox[3em]{7}&\makebox[3em]{7}&\makebox[3em]{7}&\makebox[3em]{4}&\makebox[3em]{4}&\makebox[3em]{4}&\makebox[3em]{4}&\,\,{\bf43}\\
3&\makebox[3em]{5}&\makebox[3em]{6}&\makebox[3em]{6}&\makebox[3em]{6}&\makebox[3em]{3}&\makebox[3em]{3}&\makebox[3em]{3}&\makebox[3em]{3}&\,\,{\bf35}\\
4&\makebox[3em]{4}&\makebox[3em]{4}&\makebox[3em]{4}&\makebox[3em]{4}&\makebox[3em]{2}&\makebox[3em]{2}&\makebox[3em]{2}&\makebox[3em]{2}&\,\,{\bf24}\\
5&\makebox[3em]{3}&\makebox[3em]{2}&\makebox[3em]{2}&\makebox[3em]{2}&\makebox[3em]{1}&\makebox[3em]{1}&\makebox[3em]{1}&\makebox[3em]{1}&\,\,{\bf13}\\
6&\makebox[3em]{2}&\makebox[3em]{1}&\makebox[3em]{1}&\makebox[3em]{1}&\makebox[3em]{0}&\makebox[3em]{0}&\makebox[3em]{0}&\makebox[3em]{0}&\,\,\, {\bf5}\\
7&\makebox[3em]{1}&\makebox[3em]{0}&\makebox[3em]{0}&\makebox[3em]{0}&\makebox[3em]{0}&\makebox[3em]{0}&\makebox[3em]{0}&\makebox[3em]{0}&\,\,\, {\bf1}\\
{\bf Total}&\makebox[3em]{{\bf 36}}&\makebox[3em]{{\bf 36}}&\makebox[3em]{{\bf 36}}&\makebox[3em]{{\bf 36}}&\makebox[3em]{{\bf 18}}&\makebox[3em]{{\bf 18}}&\makebox[3em]{{\bf 18}}&\makebox[3em]{{\bf 18}}&{\bf216}\\
\bottomrule
\end{tabular}
}
\caption{The number of $m_f\ge0$ hyperfine states in $^{39}$KCs spread over the various $\left| \Omega \right|$ symmetries.  
}
\label{tab:t39KCs}
\end{table}
\begin{table}[htbp]
\scalebox{0.8}{
\begin{tabular}{c| l| lll| llll| l}
\hline
\toprule
\textbf{$\Omega$} &\makebox[3em]{2}&\makebox[3em]{1}&\makebox[3em]{1$^{\alpha}$}&\makebox[3em]{1$^{\beta}$}&\makebox[3em]{0$^+_{1/2}$}&\makebox[3em]{0$^-_{1/2}$}&\makebox[3em]{0$^+_{3/2}$}&\makebox[3em]{0$^-_{3/2}$}&{\bf Tot}\\

\hline
\midrule
\textbf{$m_f$} &&&&&&&&\\
$1/2$&\makebox[3em]{13}&\makebox[3em]{15}&\makebox[3em]{15}&\makebox[3em]{15}&\makebox[3em]{8}&\makebox[3em]{8}&\makebox[3em]{8}&\makebox[3em]{8}&\,\,{\bf90}\\
$3/2$&\makebox[3em]{13}&\makebox[3em]{14}&\makebox[3em]{14}&\makebox[3em]{14}&\makebox[3em]{7}&\makebox[3em]{7}&\makebox[3em]{7}&\makebox[3em]{7}&\,\,{\bf83}\\
$5/2$&\makebox[3em]{12}&\makebox[3em]{12}&\makebox[3em]{12}&\makebox[3em]{12}&\makebox[3em]{6}&\makebox[3em]{6}&\makebox[3em]{6}&\makebox[3em]{6}&\,\,{\bf72}\\
$7/2$&\makebox[3em]{10}&\makebox[3em]{10}&\makebox[3em]{10}&\makebox[3em]{10}&\makebox[3em]{5}&\makebox[3em]{5}&\makebox[3em]{5}&\makebox[3em]{5}&\,\,{\bf60}\\
$9/2$&\makebox[3em]{8}&\makebox[3em]{8}&\makebox[3em]{8}&\makebox[3em]{8}&\makebox[3em]{4}&\makebox[3em]{4}&\makebox[3em]{4}&\makebox[3em]{4}&\,\,{\bf48}\\
$11/2$&\makebox[3em]{6}&\makebox[3em]{6}&\makebox[3em]{6}&\makebox[3em]{6}&\makebox[3em]{3}&\makebox[3em]{3}&\makebox[3em]{3}&\makebox[3em]{3}&\,\,{\bf36}\\
$13/2$&\makebox[3em]{4}&\makebox[3em]{4}&\makebox[3em]{4}&\makebox[3em]{4}&\makebox[3em]{2}&\makebox[3em]{2}&\makebox[3em]{2}&\makebox[3em]{2}&\,\,{\bf24}\\
$15/2$&\makebox[3em]{3}&\makebox[3em]{2}&\makebox[3em]{2}&\makebox[3em]{2}&\makebox[3em]{1}&\makebox[3em]{1}&\makebox[3em]{1}&\makebox[3em]{1}&\,\,{\bf13}\\
$17/2$&\makebox[3em]{2}&\makebox[3em]{1}&\makebox[3em]{1}&\makebox[3em]{1}&\makebox[3em]{0}&\makebox[3em]{0}&\makebox[3em]{0}&\makebox[3em]{0}&\,\,\, {\bf5}\\
$19/2$&\makebox[3em]{1}&\makebox[3em]{0}&\makebox[3em]{0}&\makebox[3em]{0}&\makebox[3em]{0}&\makebox[3em]{0}&\makebox[3em]{0}&\makebox[3em]{0}&\,\,\, {\bf1}\\
{\bf Total}&\makebox[3em]{{\bf 72}}&\makebox[3em]{{\bf 72}}&\makebox[3em]{{\bf 72}}&\makebox[3em]{{\bf 72}}&\makebox[3em]{{\bf 36}}&\makebox[3em]{{\bf 36}}&\makebox[3em]{{\bf 36}}&\makebox[3em]{{\bf 36}}&{\bf432}\\
\bottomrule
\end{tabular}
}
\caption{Same as Table \ref{tab:t39KCs} for $^{40}$KCs.}
\label{tab:t40KCs}
\end{table}

\section{Results}
\label{sec:results}

We present our calculations for the hfs of the first excited states of the bosonic $^{39}$KCs and fermonic $^{40}$KCs molecules, correlated to the K($4^2S_{1/2}, f_1$)+Cs($6^2P_{j_2}, f_2$) dissociation limits displayed in Fig.~\ref{fig:atomichfs}. 

\begin{figure}[h]
\includegraphics[scale=0.37]{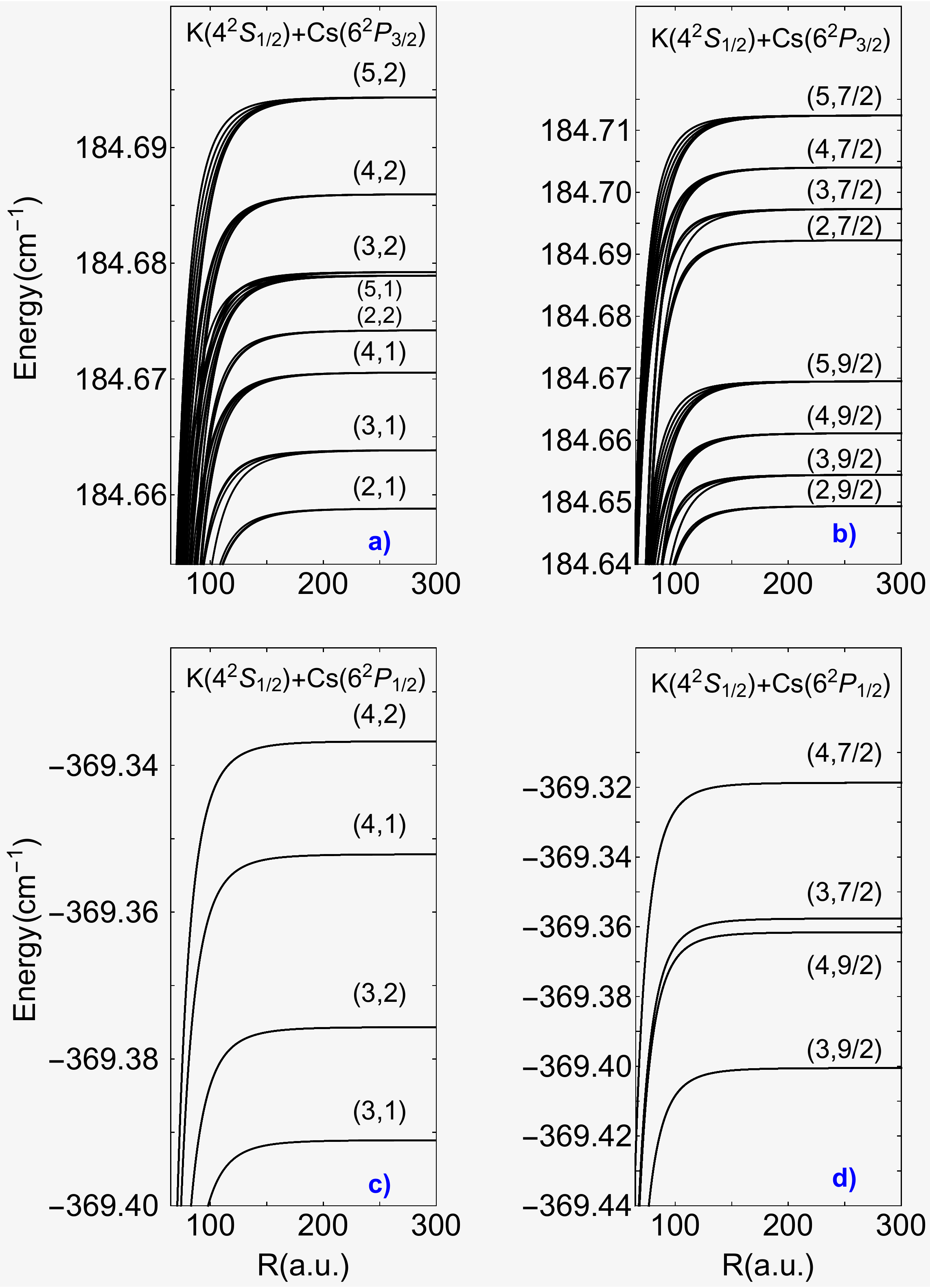}
\caption{Long-range potential energy curves (PECs) of KCs including hyperfine interaction, correlated to the K($4^2S_{1/2}, f_1$)+Cs($6^2P_{j_2}, f_2$). (a) and (c): the 216 PECs of $^{39}$KCs. (b) and (d): the 432 PECs of $^{40}$KCs. Hyperfine dissociation limits are labeled with ($f_{1}$,$f_{2}$)}.
\label{fig:hfsinset3940}
\end{figure}

The long-range PECs resulting from the diagonalization of the total Hamiltonian (Eq.~(\ref{eq:total})) in the molecular basis of Eq.~(\ref{eq:molbas}) are presented in Fig.~\ref{fig:hfsinset3940}. They are displayed down to $R \approx 100$~a.u. where the long-range van der Waals interaction is of the same magnitude than the hyperfine splittings. All PECs are attractive, as expected from Fig.~\ref{fig:hundsc}. Below 100~a.u. PECs correlated to different $(f_1,f_2)$ asymptotes strongly recouple among each other, leaving $f$ and $m_f$ as the only remaining good quantum numbers. In this range the hfs is accurately calculated and could thus be of relevance for instance for photoassociation studies. Here we are mainly interested in the hyperfine structure of deeply-bound vibrational levels relevant for STIRAP, \textit{i.e.} covering a range of lower distances.

\begin{figure}[h!]
\includegraphics[scale=0.9]{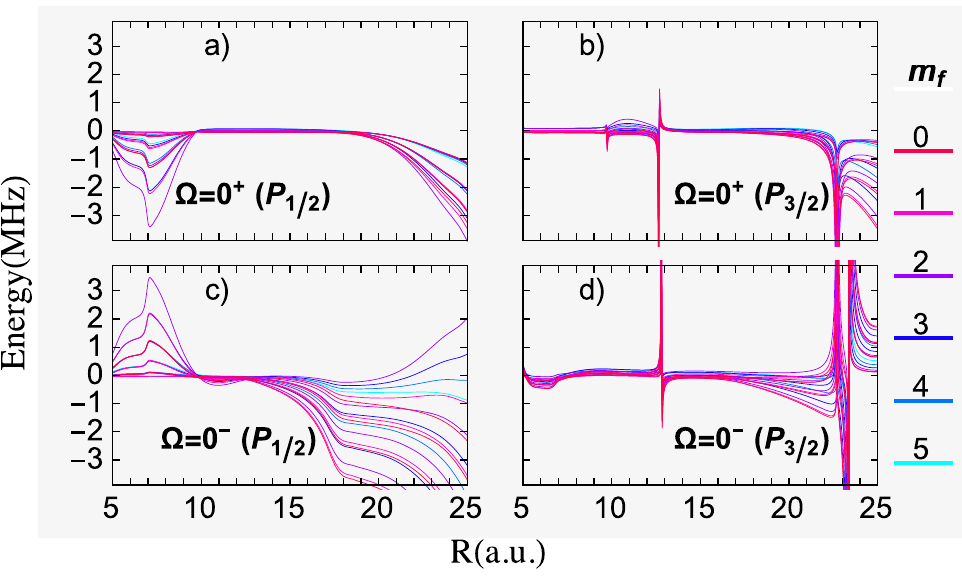}
\caption{(Color online) Difference potential energy curves for $\Omega=0$ states in $^{39}$KCs, for every value of $|m_f|$ indicated by the color code. Each panel contains 18 curves (see Table~\ref{tab:t39KCs}).}
\label{fig:omega039}
\end{figure}

\begin{figure}[h!]
\includegraphics[scale=0.9]{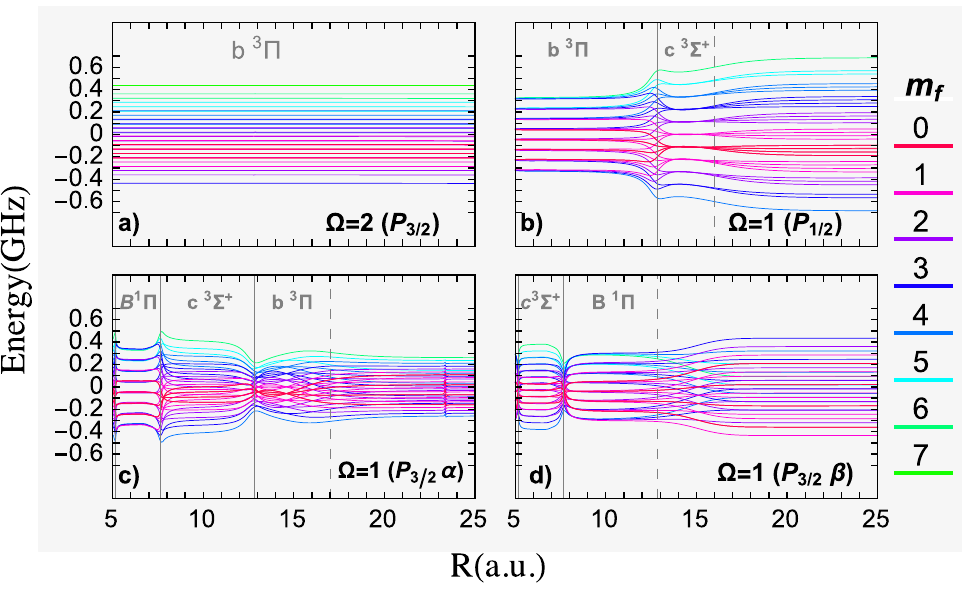}
\caption{(Color online) Difference potential energy curves for $\Omega=2, 1$ states in $^{39}$KCs, for every value of $|m_f|$ indicated by the color code. Each panel contains 36 curves (see Table~\ref{tab:t39KCs}). When appropriate, the dominant Hund's case \textit{a} state is reported at the top of the panels for each interval between two crossings of Hund's case \textit{a} potential energy curves (indicated by solid vertical lines). The dashed lines at large $R$ roughly locates the distance beyond which Hund's case \textit{a} states are strongly mixed by spin-orbit interaction.}
\label{fig:omega2139}
\end{figure}

\begin{figure}[h!]
\includegraphics[scale=0.9]{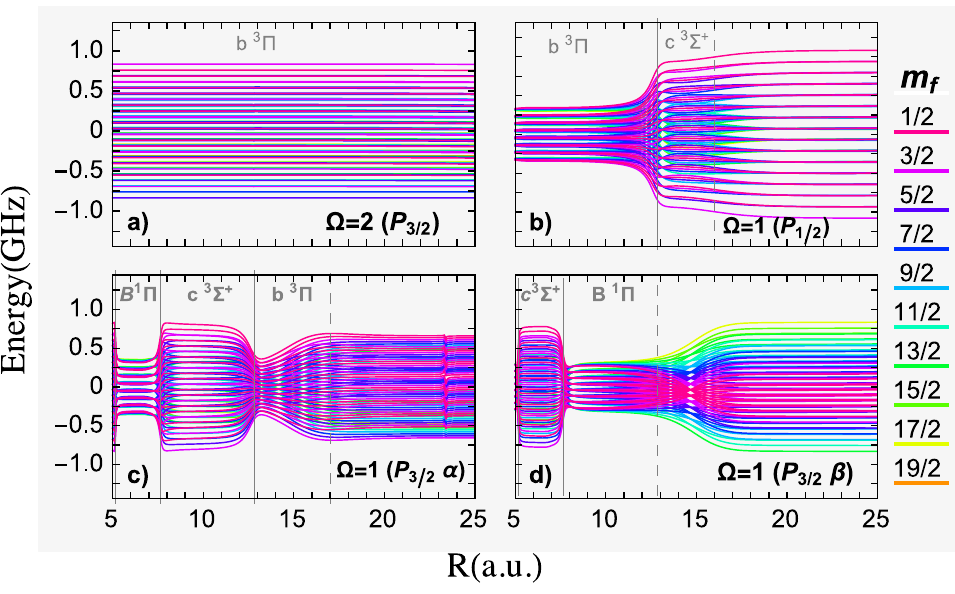}
\caption{(Color online) Same as Fig.~\ref{fig:omega2139} for $^{40}$KCs. Each panel contains 72 curves (see Table~\ref{tab:t39KCs}).} 
\label{fig:omega2140}
\end{figure}

Below an internuclear distance of about 30~a.u., the Hund's case \textit{c} PECs are well separated in energy while their hyperfine structure is much smaller than their mutual spacing. Then it is convenient to plot the difference between each of these PECs with the related pure Hund's case \textit{c} ones. The global behavior of these difference PECs (or DPECs in the following) in the range on internuclear distances between 5~a.u. and 25~a.u. is displayed for both isotopologues in Figs.~\ref{fig:omega2139} and \ref{fig:omega2140} ($\Omega=2, 1$ states) and in Figs.~\ref{fig:omega039} and \ref{fig:omega2140} ($\Omega=0$ states). For a given $\Omega$ value, the DPECs exhibit similar amplitudes of the manifold, with zones of regularities and sharp transitions between them along the $R$ interval, for both isotopologues. Most of such transitions result from the crossings between Hund's case \textit{c} PECs (see Table \ref{tab:hundsccross}) or Hund's case \textit{a} PECs (see Fig.~\ref{fig:hundsa}), which are then coupled by hf interaction so that the DPECs cannot be rigorously defined anymore. This pattern appears as more pronounced for the $\Omega=0$ states, since the hyperfine splitting of these states are with 2 to 3 orders of magnitude smaller than for $\Omega=1,2$ states. Note that the peak around 23~a.u. in the DPECs for $0^+(P_{3/2})$ and $0^- (P_{3/2})$ states, not reported in Table \ref{tab:hundsccross}, is a consequence of the closeness ($\approx0.068$~cm$^{-1}$) of these states at this location, so that their hyperfine structures overlap each other.  

\begin{table}[htbp]
\scalebox{1}{
\begin{tabular}{|c|c|c|c|}
\hline
&1 (P$_{3/2}\,{\alpha})$&0$^+$ (P$_{3/2}$)&0$^-$ (P$_{3/2}$)\\
\hline
2(P$_{3/2}$)&12.82&9.54&12.81\\
&&&\\
&&&\\
\hline
1(P$_{1/2}$)&-&{\it 9.73}&-\\
&&{\it 12.66}&\\
&&&\\
\hline
1(P$_{3/2}\,{\alpha}$)&-&-&{\it 12.80}\\
&&&{\it 23.36}\\
&&&\\
&&&\\
\hline
\end{tabular}
}\caption{Internuclear distances (in a.u.) where the KCs Hund's case {\it c} PECs cross each other. Values in italics corresponds to the location of the sharp peaks in the corresponding difference PECs for (see text).}
\label{tab:hundsccross}
\end{table}

\begin{figure}[h!]
\includegraphics[scale=0.9]{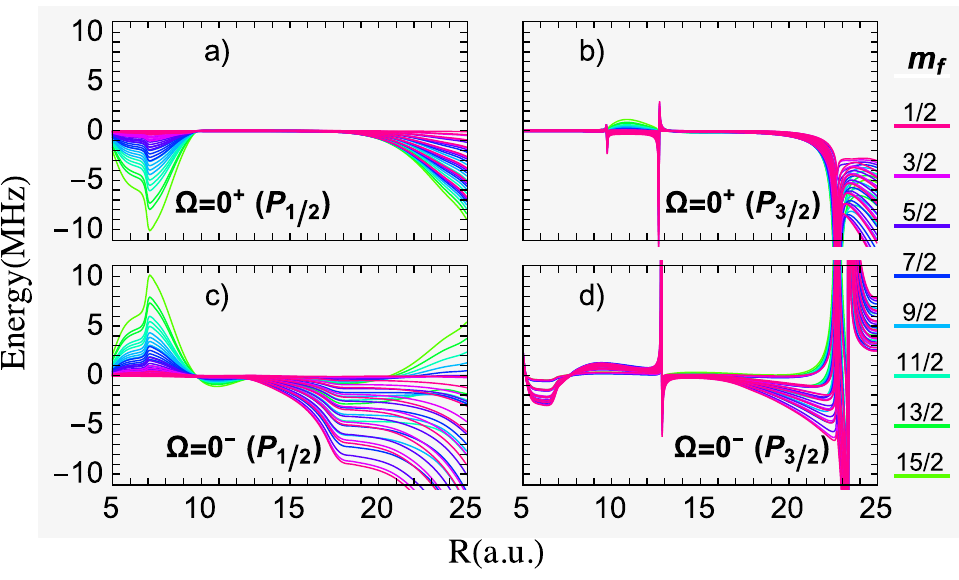}
\caption{(Color online) Difference potential energy curves for $\Omega=0$ states in $^{40}$KCs, for every value of $|m_f|$ indicated by the color code. Each panel contains 36 curves (see Table~\ref{tab:t40KCs}).}
\label{fig:omega040}
\end{figure}

In between these peaks, the DPECs may reflect the local Hund's case \textit{a} character of the hyperfine states, which is an important information for the future interpretation of the hyperfine structure of molecular levels. While our model does not explicitly treat the individual terms of the hfs Hamiltonian of Eq.\ref{eq:HFS}, one can discuss their influence by formally rewriting in an effective manner for the molecule, so that the amplitude of the hfs manifolds are easily interpreted: 
\begin{equation}
\hat{{\bf H}}^{\text{hf}}_{\text{mol}}=\hat{\bf H}^{\text{hf}}_{{\bf LI}}+ \hat{\bf H}^{\text{hf}}_{{\bf SI}}+\hat{\bf H}^{\text{hf}}_{{\bf FC}}+\hat{\bf H}^{\text{hf}}_{EQ}
\label{eq:HFS-mol}
\end{equation}
where the symbols $L$, $S$ and $I$ now refer to the angular momenta of the molecule.

The $2(P_{3/2})$ DPECs are solely built on the $b^3\Pi$ state, so that the resulting hfs is monotonous with $R$ and is spread over about 0.8~GHz for $^{39}$KCs (Fig.~\ref{fig:omega2139}a) and 1.4~GHz for $^{40}$KCs (Fig.~\ref{fig:omega2140}a). The order of the $|m_f|$ curves is contrasted between the two isotopologues: they are arranged in a somewhat disordered way in $^{40}$KCs, while they appears symmetrically around the $m_f=0$ ones in $^{39}$KCs, except for the largest $|m_f|$ values. This simple example illustrates the tight competition between the various terms of the hyperfine Hamiltonian, thus preventing from a uniform hfs pattern. In the $b^3\Pi$ state, the projection of the total electronic angular momentum as well as of the spin can not be zero. All of the three terms $\hat{\bf H}^{\text{hf}}_{{\bf LI}}$, $\hat{\bf H}^{\text{hf}}_{{\bf SI}}$, and $\hat{\bf H}^{\text{hf}}_{{\bf FC}}$ of the magnetic dipole Hamiltonian do contribute to the hfs, yielding a manifold with the largest amplitude of all the excited states.

The variations of the $1(P_{j})$ DPECs display several regimes related to the local Hund's case \textit{a} character of the states. The $|m_f|$ curves are spread symmetrically around the $m_f=0$ for the $1(P_{1/2})$ and $1(P_{3/2}\,\alpha)$ manifolds in $^{39}$KCs, and for the $1(P_{3/2}\,\beta)$ manifold in $^{40}$KCs. In contrast, the DPECs are not structured according to $|m_f|$ values for the $1(P_{1/2})$ and $1(P_{3/2}\,\alpha)$ manifolds in $^{40}$KCs, and for the $1(P_{3/2}\,\beta)$ manifold in $^{39}$KCs. The energy amplitude spanned by the DPECs strongly depends on their local character, the largest one occurring when the states are mostly of $c^3\Sigma^+$ type. The smallest hfs is found for the DPECs of $B^1\Pi$ character for which the $\hat{\bf H}^{\text{hf}}_{{\bf SI}}$ vanishes. A slightly larger hfs is visible for the DPECs of $b^3\Pi$ character for which the projection of the electronic spin on the molecular axis vanishes in $\hat{\bf H}^{\text{hf}}_{{\bf SI}}$. The largest hfs is found for the DPECs of $c^3\Sigma^+$ character as the three terms of the magnetic dipole interaction contribute to the interaction. 

\begin{figure}[h!]
\includegraphics[scale=0.9]{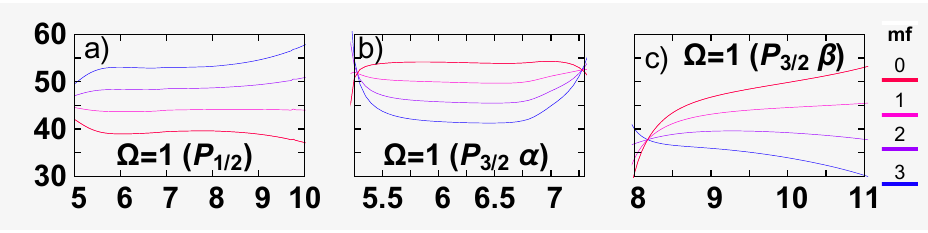}
\caption{Zooms on the $R$-intervals for which the $^{39}$KCs $1 (P_{j})$ DPECs of Fig.~\ref{fig:omega2139} are grouped according to the projection of the nuclear spin of the cesium and potassium atoms. The figure displays the four curves $m_{i_{^{39}K}}=-3/2,...,3/2$ associated to the 4th pack (counting from the top) in case of the $^{39}$K$^{133}$Cs molecule.}
\label{fig:pack39}
\end{figure}

\begin{figure}[h!]
\includegraphics[scale=0.9]{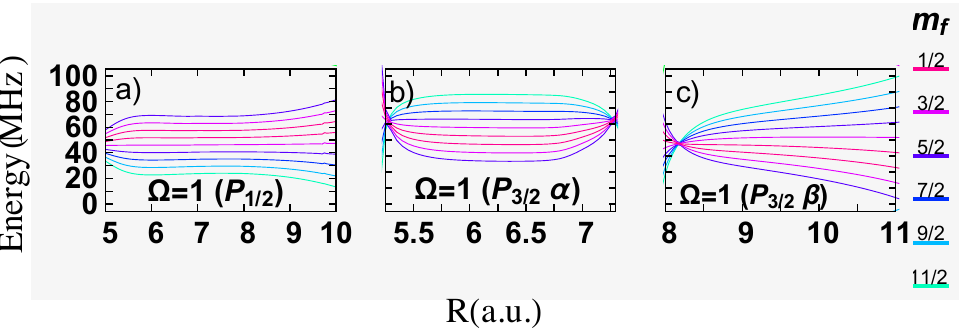}
\caption{Same as Fig.~\ref{fig:pack39} for the $^{40}$KCs $1 (P_{j})$ DPECs of Fig.~\ref{fig:omega2140}, with nine curves $m_{i_{^{40}K}}=-4,...,4$.}
\label{fig:pack40}
\end{figure}

The analysis of the structure of the DPECs for various ranges of distance reveals information on the main coupling regimes. In several cases, we observe that the DPECs are arranged in eight groups, corresponding to the possible values of the projection of the Cs nuclear spin $m_{i_{Cs}}=-7/2,...,7/2$: the $1 (P_{1/2})$ DPECs in the region with $b^3\Pi$ character (Fig.~\ref{fig:omega2139}b, Fig.~\ref{fig:omega2140}b) and the $1 (P_{3/2})$ DPECs in the region with $B^1\Pi$ character (Fig.~\ref{fig:omega2139}c, d and Fig.~\ref{fig:omega2140}c, d). Each of these eight groups exhibits a finer structure with DPECs arranged in either in four groups (Fig.~\ref{fig:pack39}) for $^{39}$KCs or nine groups (Fig.~\ref{fig:pack40}) for $^{40}$KCs molecule. This pattern correspond to the four values of $m_{i_{^{39}K}}=-3/2,...,3/2$ and to the nine values of $m_{i_{^{40}K}}=-4,...,4$, respectively. This indicates that in the related $R$ domains the nuclear spins of the two atoms do not couple each other. In $^{39}$KCs, the $1(P_{1/2})$ DPECs between 13.8~a.u. and 16~a.u. and the $1(P_{3/2}\,\alpha)$ DPECS at 14.5 a.u., mostly of $c^3\Sigma^+$ character are grouped in 11 subsets corresponding to the projections of the total nuclear spin $m_I= -5,...,5$. These subsets are equidistantly located around the $m_I=0$ one, revealing the dominant first order magnetic dipole interactions yielding energies proportional to $m_I$ \cite{broyer1978}.

The coupling of the nuclear spins appears in other cases. In $^{39}$KCs, the $1P_{1/2}$ DPECs between 13.8~a.u. and 16~a.u. and the $1(P_{3/2}\,\alpha)$ DPECS at 14.5 a.u., mostly of $c^3\Sigma^+$ character are grouped in 11 subsets corresponding to the projections of the total nuclear spin $m_I= -5,...,5$. These subsets are equidistantly located around the $m_I=0$ one, revealing the dominant first order magnetic dipole interactions yielding energies proportional to $m_I$ \cite{broyer1978}. Furthermore, Fig.~\ref{fig:omega2140}b illustrates the coupling of nuclear spins for $R>16$~a.u. in the $1(P_{1/2})$ DPECS of $^{40}$KCs with sixteen groups corresponding to the possible projections of the total nuclear spin $15/2$. This is easily understood as unlike $^{39}$KCs, the hfs of the $^{39}$K($4s$) is larger than the Cs($6p_{1/2,3/2})$ ones, so that the grouping of DPECs is different.

In contrast with the cases above, the hyperfine energy splittings for the $\Omega=0$ DPECs have a magnitude in the MHz range for both isotopologues, which can be interpreted with Eq.~\ref{eq:HFS-mol}. The first-order magnetic dipole interaction vanishes, but not the second-order magnetic dipole and higher-order multipole interactions, which give small corrections to the energy \cite{broyer1978}. 

\begin{figure}[h!]
\includegraphics[scale=0.6]{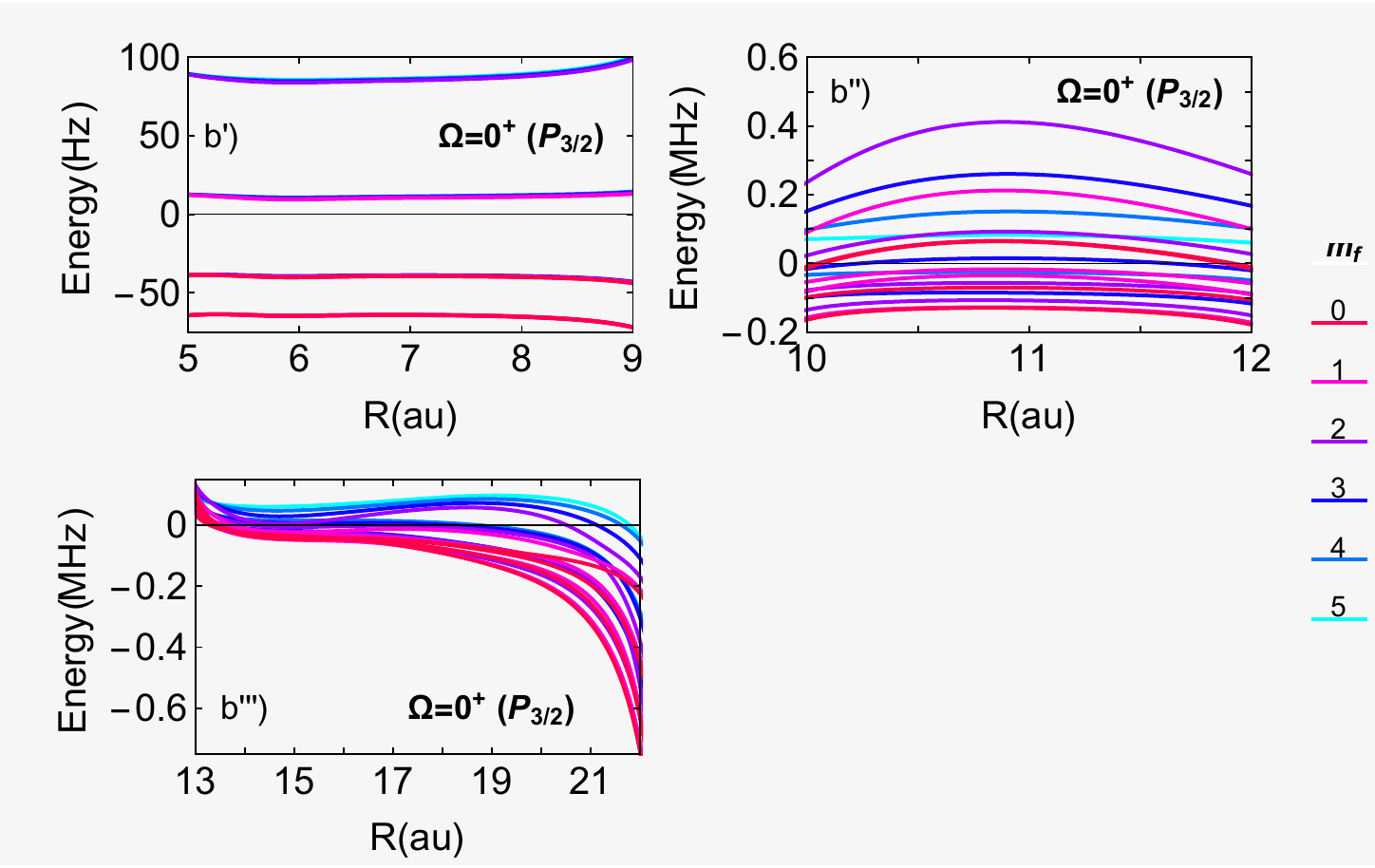}
\caption{Zooms of the $^{39}$KCs 0$^+(P_{3/2})$ DPECs of Fig.~\ref{fig:omega039}b for ranges of $R$ in between the peaks marking the transition from one Hund's case \textit{a} character to the other.}
\label{fig:omega039pz}
\end{figure}

\begin{figure}[h!]
\includegraphics[scale=0.6]{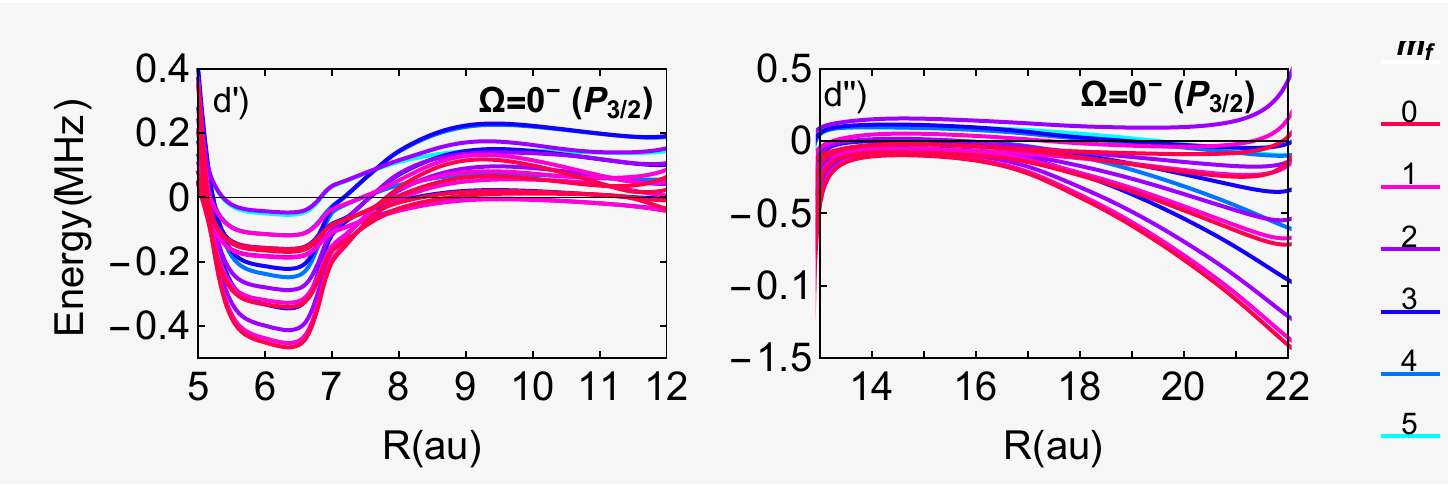}
\caption{Zooms of the $^{39}$KCs 0$^-(P_{3/2})$ DPECs of Fig.~\ref{fig:omega039}d for ranges of $R$ in between the peaks marking the transition from one Hund's case \textit{a} character to the other.}
\label{fig:omega039mz}
\end{figure}

\begin{figure}[h!]
\includegraphics[scale=0.6]{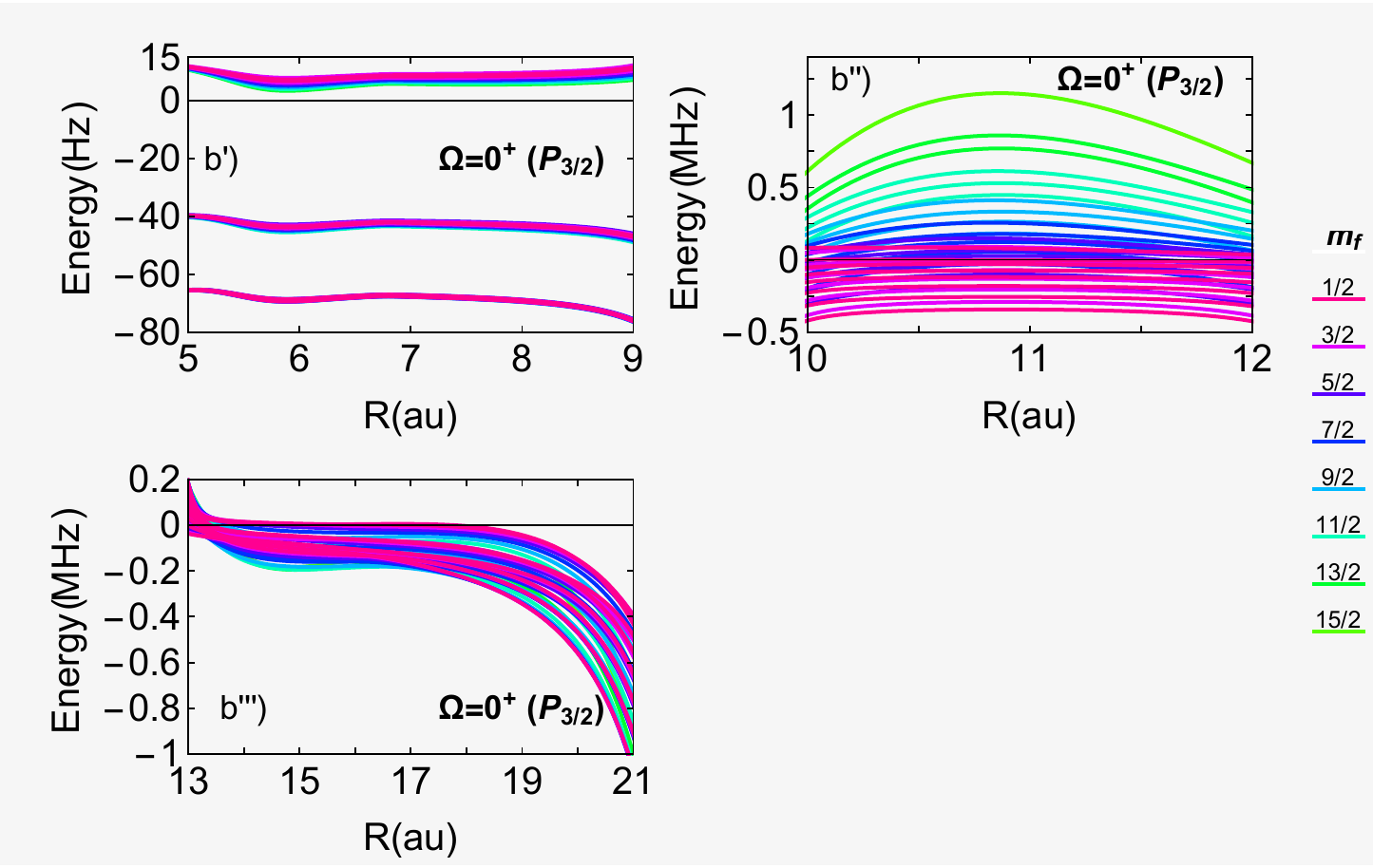}
\caption{Zooms of the $^{40}$KCs 0$^+(P_{3/2})$ DPECs of Fig.~\ref{fig:omega040}b for ranges of $R$ in between the peaks marking the transition from one Hund's case \textit{a} character to the other.}
\label{fig:omega040pz}
\end{figure}

\begin{figure}[h!]
\includegraphics[scale=0.6]{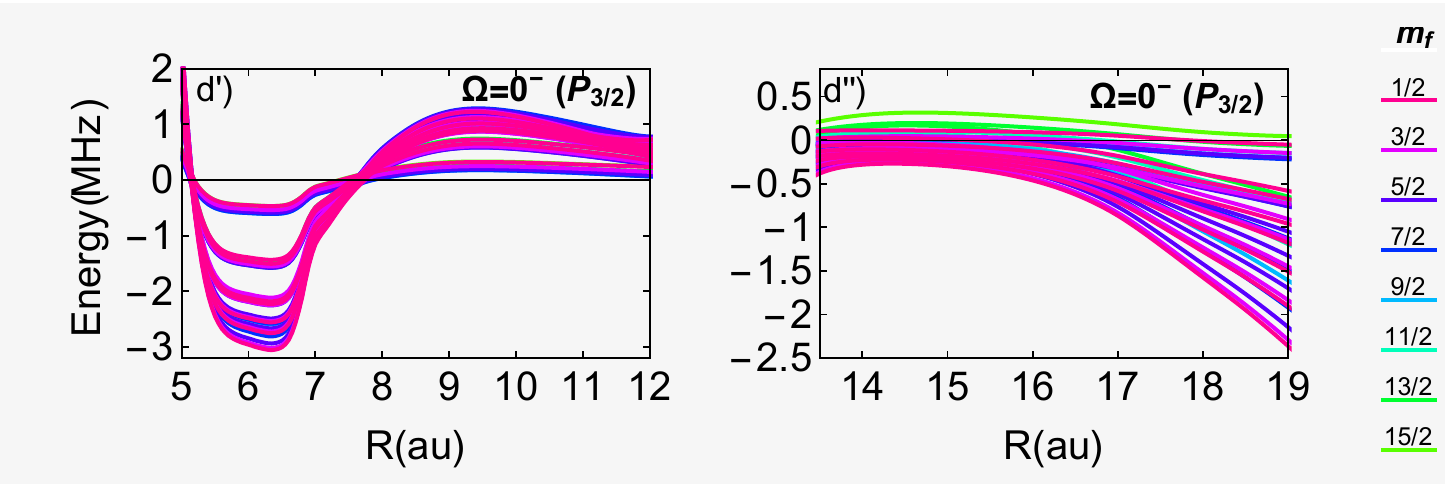}
\caption{Zooms of the $^{40}$KCs 0$^-(P_{3/2})$ DPECs of Fig.~\ref{fig:omega040}d for ranges of $R$ in between the peaks marking the transition from one Hund's case \textit{a} character to the other.}
\label{fig:omega040mz}
\end{figure}

\section{Discussion: towards a full model of the formation of ultracold KCs molecules }
\label{sec:discussion}

We have extended the asymptotic model of Ref.~\cite{comparat2000} down to small internuclear distances in order to determine the potential energy curves of electronically excited molecular states of $^{39}$KCs and $^{40}$KCs including hyperfine structure. The hyperfine splittings of the $\Omega=1,2$ molecular states have an amplitude in the GHz range, and typically three orders of magnitude less for the $\Omega=0$ states, as expected from Ref.\cite{broyer1978}. These calculations must be complemented by introducing molecular rotation and thus evaluate the hyperfine structure of the rovibrational levels of the molecules, which will be the purpose of a forthcoming paper. Nevertheless, the present calculations are already relevant for various aspects of ultracold molecule research. 

First, the implementation of \textit{optical shielding} in ultracold gases, \textit{i.e.} the suppression of inelastic collisions between ultracold atoms, relies on the modification of long-range interactions between ground-state and excited atoms by laser light \cite{weiner1999}. The experimental proof-of-principle was first achieved in trap-loss experiments induced by inelastic collisions of cold alkali atoms \cite{bali1994,sanchez-villicana1995,zilio1996,sukenik1998}. The suppression of ionizing (Penning) collisions was observed with Xe \cite{walhout1995} and Kr atoms \cite{katori1994}, as well as the suppression of reactive collisions like photoassociative collisions of sodium atoms \cite{marcassa1994}. These experiments have been achieved for gases with temperature in the 10-100~$\mu$K range, too large to allow for selecting a single repulsive channel so that the suppression effect is partial. A full description of this process at even lower temperatures must include the hyperfine structure of the related electronic states as determined in the present work. Here all the electronic states are attractive at large distances, but a few electronic states correlating to the K($4p$)+Cs($6s$) dissociation limit are indeed repulsive (see Fig.~\ref{fig:hundsa}) and thus suitable for optical shielding.

\begin{figure}[h]
\includegraphics[scale=0.35]{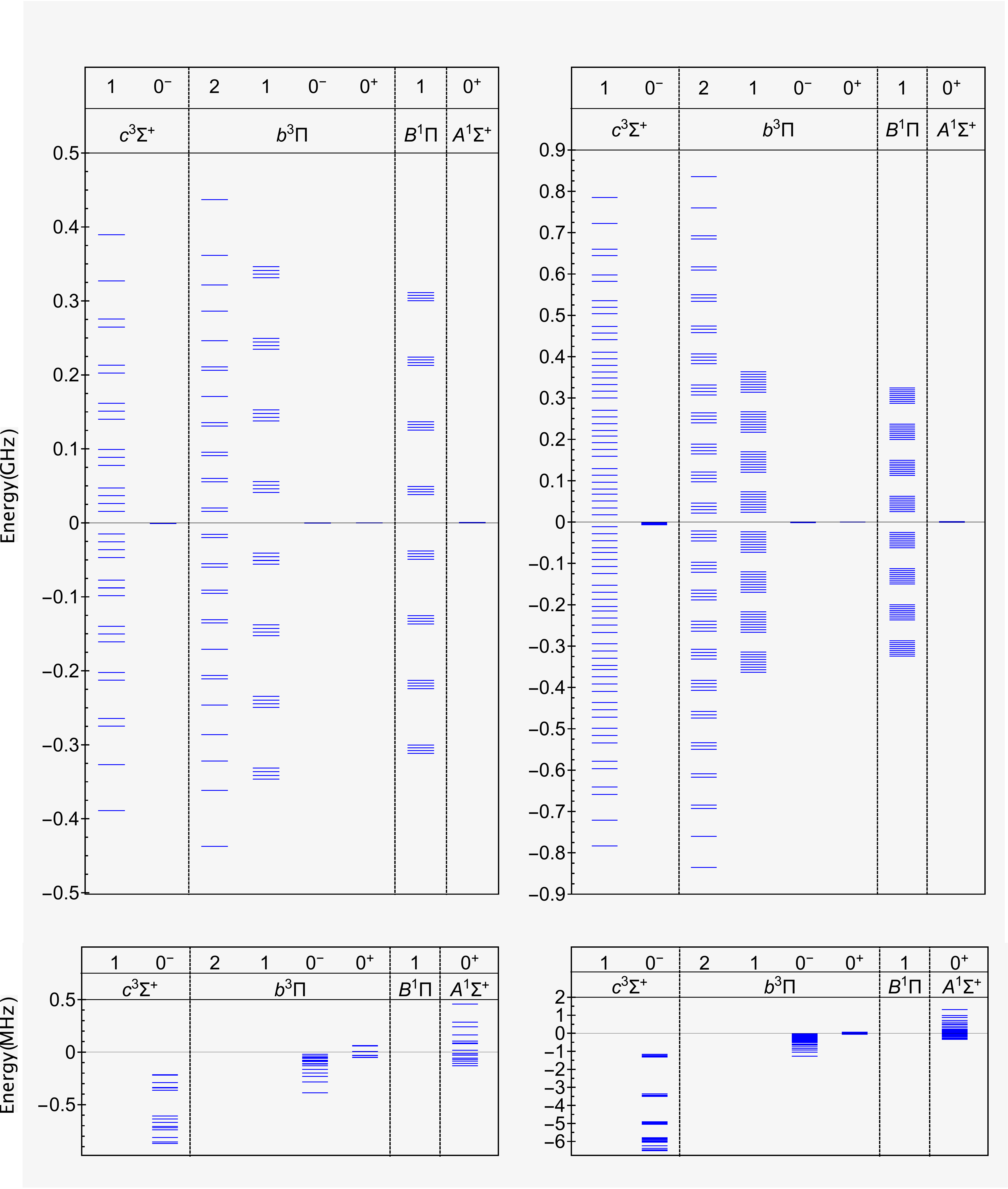}
\caption{Difference potential energies of the Hund's case \textit{c} states at $R=10.5$~a.u. of the bosonic $^{39}$KCs (left panels) and fermionic $^{40}$KCs (right panels) molecules. The lower panels display zooms for the $\Omega=0$ states. Both the relevant Hund's case \textit{c} states and the dominant  Hund's case \textit{a} character are indicated. For Hund's case \textit{c} notation in the upper line we have used a simplified one giving only the value of $\Omega$. From column 1 to column 8 the order of the states is $1(P_{3/2}\,{\alpha})$, $0^{-}(P_{3/2})$, $2(P_{3/2})$, $1(P_{1/2})$, $0^{-}(P_{1/2})$, $0^{+}(P_{1/2})$, $1(P_{3/2}{\,\beta})$,  and $0^{+}(P_{3/2})$. } 
\label{fig:HFS-10.5}
\end{figure}

Second, the most advanced achievements toward the creation of dense gases of cold molecules in their ground state rely on two main steps \cite{danzl2008,lang2008,ni2008,zirbel2008,takekoshi2014,molony2014}. First a pair of ultracold atoms is associated into a loosely-bound vibrational level of the ground molecular state manifold using magnetic or laser field. Then a controlled population transfer from this well-defined level to the absolute ground state level can be coherently performed by the method of stimulated rapid adiabatic passage (STIRAP), involving an intermediate excited molecular state. The theoretical modeling of such a multistep process without taking in account the hyperfine structure of the molecular levels yields an efficient guiding procedure for experimental work \cite{borsalino2014,borsalino2015}. However, the optimal efficiency of the transfer imposes the control of the population at the single quantum state level, namely including the hyperfine structure. The present results deliver some hints for choosing a convenient intermediate state for STIRAP. This is exemplified by Fig.~\ref{fig:HFS-10.5} showing the hyperfine structure of all the $^{39}$KCs and $^{40}$KCs PECs dissociating into K($4s\,^2S_{1/2}$)+Cs($6p\,^2P_{1/2,3/2}$) at $R_i=10.5$~a.u.. This particular distance is chosen after the theoretical study of Ref.\cite{borsalino2015}, which proposes various options for achieving STIRAP in KCs involving intermediate levels with radial wavefunctions located around $R_i$. The corresponding rotational energy estimated from the rotational constant $B=\hbar^2/(2\mu R_i^2)$ amounts for about 500~MHz. The $\Omega=0$ vibrational levels will exhibit a well-defined rotational structure with each sublevel composed of a tiny hyperfine manifold, which may be tedious to resolve experimentally. On the other hand, the $\Omega=1$ vibrational levels will be characterized by a complex structure resulting from the competition of the hyperfine interaction and the molecular rotation of comparable magnitude. As already demonstrated experimentally, such levels can be addressed individually, but their characterization in terms of quantum numbers is more involved. For an individual level, it is possible to set up an effective Hamiltonian whose parameters are fitted on the recorded spectrum \cite{takekoshi2014}. In our forthcoming work, we will model the structure of all levels starting from the atomic hyperfine parameters, and we will adjust them in order to match the observations.

\section*{Appendix}

We present in Table~\ref{tab:HFS-atoms} the values of the hyperfine energy splittings of the $^{39}$K and $^{40}$K $4s$ level, and of the Cs $6p\,^2P_{1/2,3/2}$ levels in terms of the related molecular dissociation limits.

\begin{table}[h]
\begin{tabular}{|c|c|c||c||c||}
\hline
\toprule
&&&&\\
\textbf{K({\it 4s})+Cs({\it 6p})} & 
\textbf{K({\it 4s}$_{j_K}$)+Cs({\it 6p}$_{j_{Cs}}$)}&
&
\textbf{$^{39}$K({\it 4s}$_{j_K}$)+$^{133}$Cs({\it 6p}$_{j_{Cs}}$)}&
\textbf{$^{40}$K({\it 4s}$_{j_K}$)+$^{133}$Cs({\it 6p}$_{j_{Cs}}$)}
\\
(l$_K$,l$_{Cs}$)&
(j$_K$,j$_{Cs}$)&
E$_{f}$(cm$^{-1}$)&
E$_{hf}$(cm$^{-1}$)\,\,\,\,\,\,\,\,\,\,(f$_{Cs}$,f$_{K}$)&
E$_{hf}$(cm$^{-1}$)\,\,\,\,\,\,\,\,\,\,(f$_{Cs}$,f$_{K}$)
\\
&&&&\\
\hline
&&&184.7498\,\,\,\,\,\,\,\,\,\,\,\,\,\,\,\,\,\,(5,\,2)&184.712\,\,\,\,\,\,\,\,\,\,\,\,\,\,\,\,\,\,(5,\,7/2)\\
 \cline{4-5}
&&&184.7415\,\,\,\,\,\,\,\,\,\,\,\,\,\,\,\,\,\,(4,\,2)&184.704\,\,\,\,\,\,\,\,\,\,\,\,\,\,\,\,\,\,(4,\,7/2)\\
 \cline{4-5}
 &&&184.7345\,\,\,\,\,\,\,\,\,\,\,\,\,\,\,\,\,\,(3,\,2)&184.697\,\,\,\,\,\,\,\,\,\,\,\,\,\,\,\,\,\,(3,\,7/2)\\
 \cline{4-5}
&(1/2,\,3/2)&184.7351&184.7341\,\,\,\,\,\,\,\,\,\,\,\,\,\,\,\,\,\,(5,\,1)&184.692\,\,\,\,\,\,\,\,\,\,\,\,\,\,\,\,\,\,(2,\,7/2)\\
 \cline{4-5}
&&&184.7294\,\,\,\,\,\,\,\,\,\,\,\,\,\,\,\,\,\,\,(2,\,2)&184.67\,\,\,\,\,\,\,\,\,\,\,\,\,\,\,\,\,\,\,\,\,(5,\,9/2)\\
 \cline{4-5}
 &&&184.7258\,\,\,\,\,\,\,\,\,\,\,\,\,\,\,\,\,\,(4,\,1)&184.661\,\,\,\,\,\,\,\,\,\,\,\,\,\,\,\,\,\,(4,\,9/2)\\
 \cline{4-5}
  &&&184.7191\,\,\,\,\,\,\,\,\,\,\,\,\,\,\,\,\,\,(3,\,1)&184.654\,\,\,\,\,\,\,\,\,\,\,\,\,\,\,\,\,\,(3,\,9/2)\\
 \cline{4-5}
 (0,\,1) &&&184.7141\,\,\,\,\,\,\,\,\,\,\,\,\,\,\,\,\,\,(2,\,1)&184.649\,\,\,\,\,\,\,\,\,\,\,\,\,\,\,\,\,\,(2,\,9/2)\\
 \cline{2-5}
  &&&-369.4463\,\,\,\,\,\,\,\,\,\,\,\,\,\,\,\,\,\,(4,\,2)&-369.319\,\,\,\,\,\,\,\,\,\,\,\,\,\,\,\,\,\,(4,\,7/2)\\
 \cline{4-5}
   &(1/2,\,1/2)&-369.4696&-369.4629\,\,\,\,\,\,\,\,\,\,\,\,\,\,\,\,\,\,(4,\,1)&-369.358\,\,\,\,\,\,\,\,\,\,\,\,\,\,\,\,\,\,(3,\,7/2)\\
 \cline{4-5}
   &&&-369.4863\,\,\,\,\,\,\,\,\,\,\,\,\,\,\,\,\,\,(3,\,2)&-369.362\,\,\,\,\,\,\,\,\,\,\,\,\,\,\,\,\,\,(4,\,9/2)\\
 \cline{4-5}
   &&&-369.5030\,\,\,\,\,\,\,\,\,\,\,\,\,\,\,\,\,\,(3,\,1)&-369.401\,\,\,\,\,\,\,\,\,\,\,\,\,\,\,\,\,\,(3,\,9/2)\\
 \hline
\midrule

\bottomrule
\end{tabular}
\caption{Asymptotic molecular hyperfine energies of the $^{39}$K$^{133}$Cs and $^{40}$K$^{133}$Cs molecules for the K($4s\,^2S_{1/2}$)+Cs($6p\,^2P_{1/2,3/2}$) dissociation limits, see fig. \ref{fig:atomichfs}. }
\label{tab:HFS-atoms}
\end{table}

\section*{Acknowledgments}
A.O. acknowledges partial support from \textit{Insitut Francilien de Recherches sur les Atomes Froids} (IFRAF), and from \textit{Agence Nationale de la Recherche} (ANR), under the project COPOMOL (contract ANR-13-IS04-0004-01) and BLUESHIELD (contract ANR-14-CE34-0006-01). H.-C. N. gratefully acknowledges funding by the European Research Council (ERC) under Project No. 278417 and by the Austrian Science Foundation (FWF) under Project No. P1789-N20 (as part of BLUESHIELD).

\newpage


\begin{thebibliography}{86}
\expandafter\ifx\csname natexlab\endcsname\relax\def\natexlab#1{#1}\fi
\expandafter\ifx\csname bibnamefont\endcsname\relax
  \def\bibnamefont#1{#1}\fi
\expandafter\ifx\csname bibfnamefont\endcsname\relax
  \def\bibfnamefont#1{#1}\fi
\expandafter\ifx\csname citenamefont\endcsname\relax
  \def\citenamefont#1{#1}\fi
\expandafter\ifx\csname url\endcsname\relax
  \def\url#1{\texttt{#1}}\fi
\expandafter\ifx\csname urlprefix\endcsname\relax\def\urlprefix{URL }\fi
\providecommand{\bibinfo}[2]{#2}
\providecommand{\eprint}[2][]{\url{#2}}

\bibitem[{\citenamefont{Anderson et~al.}(1995)\citenamefont{Anderson, Ensher,
  Matthews, Wieman, and Cornell}}]{anderson1995}
\bibinfo{author}{\bibfnamefont{M.}~\bibnamefont{Anderson}},
  \bibinfo{author}{\bibfnamefont{J.}~\bibnamefont{Ensher}},
  \bibinfo{author}{\bibfnamefont{M.}~\bibnamefont{Matthews}},
  \bibinfo{author}{\bibfnamefont{C.}~\bibnamefont{Wieman}}, \bibnamefont{and}
  \bibinfo{author}{\bibfnamefont{E.}~\bibnamefont{Cornell}},
  \bibinfo{journal}{Science} \textbf{\bibinfo{volume}{269}},
  \bibinfo{pages}{198} (\bibinfo{year}{1995}).

\bibitem[{\citenamefont{Davis et~al.}(1995)\citenamefont{Davis, Mewes, Andrews,
  van Druten, Durfee, Kurn, and Ketterle}}]{davis1995}
\bibinfo{author}{\bibfnamefont{K.~B.} \bibnamefont{Davis}},
  \bibinfo{author}{\bibfnamefont{M.~O.} \bibnamefont{Mewes}},
  \bibinfo{author}{\bibfnamefont{M.~R.} \bibnamefont{Andrews}},
  \bibinfo{author}{\bibfnamefont{N.~J.} \bibnamefont{van Druten}},
  \bibinfo{author}{\bibfnamefont{D.~S.} \bibnamefont{Durfee}},
  \bibinfo{author}{\bibfnamefont{D.~M.} \bibnamefont{Kurn}}, \bibnamefont{and}
  \bibinfo{author}{\bibfnamefont{W.}~\bibnamefont{Ketterle}},
  \bibinfo{journal}{Phys. Rev. Lett.} \textbf{\bibinfo{volume}{75}},
  \bibinfo{pages}{3969} (\bibinfo{year}{1995}).

\bibitem[{\citenamefont{Bradley et~al.}(1995)\citenamefont{Bradley, Sackett,
  Tollett, and Hulet}}]{bradley1995}
\bibinfo{author}{\bibfnamefont{C.}~\bibnamefont{Bradley}},
  \bibinfo{author}{\bibfnamefont{C.}~\bibnamefont{Sackett}},
  \bibinfo{author}{\bibfnamefont{J.}~\bibnamefont{Tollett}}, \bibnamefont{and}
  \bibinfo{author}{\bibfnamefont{R.}~\bibnamefont{Hulet}},
  \bibinfo{journal}{Phys. Rev. Lett.} \textbf{\bibinfo{volume}{75}},
  \bibinfo{pages}{1687} (\bibinfo{year}{1995}).

\bibitem[{\citenamefont{DeMarco and Jin}(1999)}]{demarco1999}
\bibinfo{author}{\bibfnamefont{B.}~\bibnamefont{DeMarco}} \bibnamefont{and}
  \bibinfo{author}{\bibfnamefont{D.~S.} \bibnamefont{Jin}},
  \bibinfo{journal}{Science} \textbf{\bibinfo{volume}{285}},
  \bibinfo{pages}{1703} (\bibinfo{year}{1999}).

\bibitem[{\citenamefont{Heinzen et~al.}(2000)\citenamefont{Heinzen, Wynar,
  Drummond, and Kheruntsyan}}]{heinzen2000}
\bibinfo{author}{\bibfnamefont{D.~J.} \bibnamefont{Heinzen}},
  \bibinfo{author}{\bibfnamefont{R.}~\bibnamefont{Wynar}},
  \bibinfo{author}{\bibfnamefont{P.~D.} \bibnamefont{Drummond}},
  \bibnamefont{and} \bibinfo{author}{\bibfnamefont{K.~V.}
  \bibnamefont{Kheruntsyan}}, \bibinfo{journal}{Phys. Rev. Lett.}
  \textbf{\bibinfo{volume}{84}}, \bibinfo{pages}{5029} (\bibinfo{year}{2000}).

\bibitem[{\citenamefont{Donley et~al.}(2002)\citenamefont{Donley, Claussen,
  Thompson, and Wieman}}]{donley2002}
\bibinfo{author}{\bibfnamefont{E.~A.} \bibnamefont{Donley}},
  \bibinfo{author}{\bibfnamefont{N.~R.} \bibnamefont{Claussen}},
  \bibinfo{author}{\bibfnamefont{S.~T.} \bibnamefont{Thompson}},
  \bibnamefont{and} \bibinfo{author}{\bibfnamefont{C.~E.}
  \bibnamefont{Wieman}}, \bibinfo{journal}{Nature}
  \textbf{\bibinfo{volume}{417}}, \bibinfo{pages}{529} (\bibinfo{year}{2002}).

\bibitem[{\citenamefont{Jochim et~al.}(2003)\citenamefont{Jochim, Bartenstein,
  Altmeyer, Hendl, Riedl, Chin, Denschlag, and Grimm}}]{jochim2003a}
\bibinfo{author}{\bibfnamefont{S.}~\bibnamefont{Jochim}},
  \bibinfo{author}{\bibfnamefont{M.}~\bibnamefont{Bartenstein}},
  \bibinfo{author}{\bibfnamefont{A.}~\bibnamefont{Altmeyer}},
  \bibinfo{author}{\bibfnamefont{G.}~\bibnamefont{Hendl}},
  \bibinfo{author}{\bibfnamefont{S.}~\bibnamefont{Riedl}},
  \bibinfo{author}{\bibfnamefont{C.}~\bibnamefont{Chin}},
  \bibinfo{author}{\bibfnamefont{J.~H.} \bibnamefont{Denschlag}},
  \bibnamefont{and} \bibinfo{author}{\bibfnamefont{R.}~\bibnamefont{Grimm}},
  \bibinfo{journal}{Science} \textbf{\bibinfo{volume}{302}},
  \bibinfo{pages}{2101} (\bibinfo{year}{2003}).

\bibitem[{\citenamefont{Herbig et~al.}(2003)\citenamefont{Herbig, Kraemer,
  Mark, Weber, Chin, N\"{a}gerl, and Grimm}}]{herbig2003}
\bibinfo{author}{\bibfnamefont{J.}~\bibnamefont{Herbig}},
  \bibinfo{author}{\bibfnamefont{T.}~\bibnamefont{Kraemer}},
  \bibinfo{author}{\bibfnamefont{M.}~\bibnamefont{Mark}},
  \bibinfo{author}{\bibfnamefont{T.}~\bibnamefont{Weber}},
  \bibinfo{author}{\bibfnamefont{C.}~\bibnamefont{Chin}},
  \bibinfo{author}{\bibfnamefont{H.-C.} \bibnamefont{N\"{a}gerl}},
  \bibnamefont{and} \bibinfo{author}{\bibfnamefont{R.}~\bibnamefont{Grimm}},
  \bibinfo{journal}{Science} \textbf{\bibinfo{volume}{301}},
  \bibinfo{pages}{1510} (\bibinfo{year}{2003}).

\bibitem[{\citenamefont{Zwierlein et~al.}(2003)\citenamefont{Zwierlein, Stan,
  Schunck, Raupach, Gupta, Hadzibabic, and Ketterle}}]{zwierlein2003}
\bibinfo{author}{\bibfnamefont{M.~W.} \bibnamefont{Zwierlein}},
  \bibinfo{author}{\bibfnamefont{C.~A.} \bibnamefont{Stan}},
  \bibinfo{author}{\bibfnamefont{C.~H.} \bibnamefont{Schunck}},
  \bibinfo{author}{\bibfnamefont{S.~M.~F.} \bibnamefont{Raupach}},
  \bibinfo{author}{\bibfnamefont{S.}~\bibnamefont{Gupta}},
  \bibinfo{author}{\bibfnamefont{Z.}~\bibnamefont{Hadzibabic}},
  \bibnamefont{and} \bibinfo{author}{\bibfnamefont{W.}~\bibnamefont{Ketterle}},
  \bibinfo{journal}{Phys. Rev. Lett.} \textbf{\bibinfo{volume}{91}},
  \bibinfo{pages}{250401} (\bibinfo{year}{2003}).

\bibitem[{\citenamefont{Greiner et~al.}(2003)\citenamefont{Greiner, Regal, and
  Jin}}]{greiner2003}
\bibinfo{author}{\bibfnamefont{M.}~\bibnamefont{Greiner}},
  \bibinfo{author}{\bibfnamefont{C.~A.} \bibnamefont{Regal}}, \bibnamefont{and}
  \bibinfo{author}{\bibfnamefont{D.~S.} \bibnamefont{Jin}},
  \bibinfo{journal}{Nature} \textbf{\bibinfo{volume}{426}},
  \bibinfo{pages}{537} (\bibinfo{year}{2003}).

\bibitem[{\citenamefont{Ni et~al.}(2008)\citenamefont{Ni, Ospelkaus,
  de~Miranda, Peer, Neyenhuis, Zirbel, Kotochigova, Julienne, Jin, and
  Ye}}]{ni2008}
\bibinfo{author}{\bibfnamefont{K.-K.} \bibnamefont{Ni}},
  \bibinfo{author}{\bibfnamefont{S.}~\bibnamefont{Ospelkaus}},
  \bibinfo{author}{\bibfnamefont{M.~H.~G.} \bibnamefont{de~Miranda}},
  \bibinfo{author}{\bibfnamefont{A.}~\bibnamefont{Peer}},
  \bibinfo{author}{\bibfnamefont{B.}~\bibnamefont{Neyenhuis}},
  \bibinfo{author}{\bibfnamefont{J.~J.} \bibnamefont{Zirbel}},
  \bibinfo{author}{\bibfnamefont{S.}~\bibnamefont{Kotochigova}},
  \bibinfo{author}{\bibfnamefont{P.~S.} \bibnamefont{Julienne}},
  \bibinfo{author}{\bibfnamefont{D.~S.} \bibnamefont{Jin}}, \bibnamefont{and}
  \bibinfo{author}{\bibfnamefont{J.}~\bibnamefont{Ye}},
  \bibinfo{journal}{Science} \textbf{\bibinfo{volume}{322}},
  \bibinfo{pages}{231} (\bibinfo{year}{2008}).

\bibitem[{\citenamefont{Danzl et~al.}(2008)\citenamefont{Danzl, Haller,
  Gustavsson, Mark, Hart, Bouloufa, Dulieu, Ritsch, and
  N\"{a}gerl}}]{danzl2008}
\bibinfo{author}{\bibfnamefont{J.~G.} \bibnamefont{Danzl}},
  \bibinfo{author}{\bibfnamefont{E.}~\bibnamefont{Haller}},
  \bibinfo{author}{\bibfnamefont{M.}~\bibnamefont{Gustavsson}},
  \bibinfo{author}{\bibfnamefont{M.~J.} \bibnamefont{Mark}},
  \bibinfo{author}{\bibfnamefont{R.}~\bibnamefont{Hart}},
  \bibinfo{author}{\bibfnamefont{N.}~\bibnamefont{Bouloufa}},
  \bibinfo{author}{\bibfnamefont{O.}~\bibnamefont{Dulieu}},
  \bibinfo{author}{\bibfnamefont{H.}~\bibnamefont{Ritsch}}, \bibnamefont{and}
  \bibinfo{author}{\bibfnamefont{H.-C.} \bibnamefont{N\"{a}gerl}},
  \bibinfo{journal}{Science} \textbf{\bibinfo{volume}{321}},
  \bibinfo{pages}{1062} (\bibinfo{year}{2008}).

\bibitem[{\citenamefont{Lang et~al.}(2008{\natexlab{a}})\citenamefont{Lang,
  Winkler, Strauss, Grimm, and Denschlag}}]{lang2008a}
\bibinfo{author}{\bibfnamefont{F.}~\bibnamefont{Lang}},
  \bibinfo{author}{\bibfnamefont{K.}~\bibnamefont{Winkler}},
  \bibinfo{author}{\bibfnamefont{C.}~\bibnamefont{Strauss}},
  \bibinfo{author}{\bibfnamefont{R.}~\bibnamefont{Grimm}}, \bibnamefont{and}
  \bibinfo{author}{\bibfnamefont{J.~H.} \bibnamefont{Denschlag}},
  \bibinfo{journal}{Phys. Rev. Lett.} \textbf{\bibinfo{volume}{101}},
  \bibinfo{pages}{133005} (\bibinfo{year}{2008}{\natexlab{a}}).

\bibitem[{\citenamefont{Danzl et~al.}(2010)\citenamefont{Danzl, Mark, Haller,
  Gustavsson, Hart, Aldegunde, Hutson, and N\"agerl}}]{danzl2010}
\bibinfo{author}{\bibfnamefont{J.~G.} \bibnamefont{Danzl}},
  \bibinfo{author}{\bibfnamefont{M.~J.} \bibnamefont{Mark}},
  \bibinfo{author}{\bibfnamefont{E.}~\bibnamefont{Haller}},
  \bibinfo{author}{\bibfnamefont{M.}~\bibnamefont{Gustavsson}},
  \bibinfo{author}{\bibfnamefont{R.}~\bibnamefont{Hart}},
  \bibinfo{author}{\bibfnamefont{J.}~\bibnamefont{Aldegunde}},
  \bibinfo{author}{\bibfnamefont{J.~M.} \bibnamefont{Hutson}},
  \bibnamefont{and} \bibinfo{author}{\bibfnamefont{H.-C.}
  \bibnamefont{N\"agerl}}, \bibinfo{journal}{Nature Phys.}
  \textbf{\bibinfo{volume}{6}}, \bibinfo{pages}{265} (\bibinfo{year}{2010}).

\bibitem[{\citenamefont{Takekoshi et~al.}(2014)\citenamefont{Takekoshi,
  Reichs\"ollner, Schindewolf, Hutson, Sueur, Dulieu, Ferlaino, Grimm, and
  N\"agerl}}]{takekoshi2014}
\bibinfo{author}{\bibfnamefont{T.}~\bibnamefont{Takekoshi}},
  \bibinfo{author}{\bibfnamefont{L.}~\bibnamefont{Reichs\"ollner}},
  \bibinfo{author}{\bibfnamefont{A.}~\bibnamefont{Schindewolf}},
  \bibinfo{author}{\bibfnamefont{J.~M.} \bibnamefont{Hutson}},
  \bibinfo{author}{\bibfnamefont{C.~R.~L.} \bibnamefont{Sueur}},
  \bibinfo{author}{\bibfnamefont{O.}~\bibnamefont{Dulieu}},
  \bibinfo{author}{\bibfnamefont{F.}~\bibnamefont{Ferlaino}},
  \bibinfo{author}{\bibfnamefont{R.}~\bibnamefont{Grimm}}, \bibnamefont{and}
  \bibinfo{author}{\bibfnamefont{H.-C.} \bibnamefont{N\"agerl}},
  \bibinfo{journal}{Phys. Rev. Lett.} \textbf{\bibinfo{volume}{113}},
  \bibinfo{pages}{205301} (\bibinfo{year}{2014}).

\bibitem[{\citenamefont{Molony et~al.}(2014)\citenamefont{Molony, Gregory, Ji,
  Lu, K\"oppinger, Le~Sueur, Blackley, Hutson, and Cornish}}]{molony2014}
\bibinfo{author}{\bibfnamefont{P.~K.} \bibnamefont{Molony}},
  \bibinfo{author}{\bibfnamefont{P.~D.} \bibnamefont{Gregory}},
  \bibinfo{author}{\bibfnamefont{Z.}~\bibnamefont{Ji}},
  \bibinfo{author}{\bibfnamefont{B.}~\bibnamefont{Lu}},
  \bibinfo{author}{\bibfnamefont{M.~P.} \bibnamefont{K\"oppinger}},
  \bibinfo{author}{\bibfnamefont{C.~R.} \bibnamefont{Le~Sueur}},
  \bibinfo{author}{\bibfnamefont{C.~L.} \bibnamefont{Blackley}},
  \bibinfo{author}{\bibfnamefont{J.~M.} \bibnamefont{Hutson}},
  \bibnamefont{and} \bibinfo{author}{\bibfnamefont{S.~L.}
  \bibnamefont{Cornish}}, \bibinfo{journal}{Phys. Rev. Lett.}
  \textbf{\bibinfo{volume}{113}}, \bibinfo{pages}{255301}
  (\bibinfo{year}{2014}).

\bibitem[{\citenamefont{K\"ohler et~al.}(2006)\citenamefont{K\"ohler,
  G\'{o}ral, and Julienne}}]{kohler2006}
\bibinfo{author}{\bibfnamefont{T.}~\bibnamefont{K\"ohler}},
  \bibinfo{author}{\bibfnamefont{K.}~\bibnamefont{G\'{o}ral}},
  \bibnamefont{and} \bibinfo{author}{\bibfnamefont{P.~S.}
  \bibnamefont{Julienne}}, \bibinfo{journal}{Rev. Mod. Phys.}
  \textbf{\bibinfo{volume}{78}}, \bibinfo{pages}{1311} (\bibinfo{year}{2006}).

\bibitem[{\citenamefont{Chin et~al.}(2010)\citenamefont{Chin, Grimm, Julienne,
  and Tiesinga}}]{chin2010}
\bibinfo{author}{\bibfnamefont{C.}~\bibnamefont{Chin}},
  \bibinfo{author}{\bibfnamefont{R.}~\bibnamefont{Grimm}},
  \bibinfo{author}{\bibfnamefont{P.}~\bibnamefont{Julienne}}, \bibnamefont{and}
  \bibinfo{author}{\bibfnamefont{E.}~\bibnamefont{Tiesinga}},
  \bibinfo{journal}{Rev. Mod. Phys.} \textbf{\bibinfo{volume}{82}},
  \bibinfo{pages}{1225} (\bibinfo{year}{2010}).

\bibitem[{\citenamefont{Mark et~al.}(2007)\citenamefont{Mark, Kraemer,
  Waldburger, Herbig, Chin, N\"agerl, and Grimm}}]{mark2007b}
\bibinfo{author}{\bibfnamefont{M.}~\bibnamefont{Mark}},
  \bibinfo{author}{\bibfnamefont{T.}~\bibnamefont{Kraemer}},
  \bibinfo{author}{\bibfnamefont{P.}~\bibnamefont{Waldburger}},
  \bibinfo{author}{\bibfnamefont{J.}~\bibnamefont{Herbig}},
  \bibinfo{author}{\bibfnamefont{C.}~\bibnamefont{Chin}},
  \bibinfo{author}{\bibfnamefont{H.-C.} \bibnamefont{N\"agerl}},
  \bibnamefont{and} \bibinfo{author}{\bibfnamefont{R.}~\bibnamefont{Grimm}},
  \bibinfo{journal}{Phys. Rev. Lett.} \textbf{\bibinfo{volume}{99}},
  \bibinfo{pages}{113201} (\bibinfo{year}{2007}).

\bibitem[{\citenamefont{Lang et~al.}(2008{\natexlab{b}})\citenamefont{Lang,
  van~der Straten, Brandst\"{a}tter, Thalhammer, Winkler, Julienne, Grimm, and
  Denschlag}}]{lang2008}
\bibinfo{author}{\bibfnamefont{F.}~\bibnamefont{Lang}},
  \bibinfo{author}{\bibfnamefont{P.}~\bibnamefont{van~der Straten}},
  \bibinfo{author}{\bibfnamefont{B.}~\bibnamefont{Brandst\"{a}tter}},
  \bibinfo{author}{\bibfnamefont{G.}~\bibnamefont{Thalhammer}},
  \bibinfo{author}{\bibfnamefont{K.}~\bibnamefont{Winkler}},
  \bibinfo{author}{\bibfnamefont{P.~S.} \bibnamefont{Julienne}},
  \bibinfo{author}{\bibfnamefont{R.}~\bibnamefont{Grimm}}, \bibnamefont{and}
  \bibinfo{author}{\bibfnamefont{J.~H.} \bibnamefont{Denschlag}},
  \bibinfo{journal}{Nature Physics} \textbf{\bibinfo{volume}{4}},
  \bibinfo{pages}{223} (\bibinfo{year}{2008}{\natexlab{b}}).

\bibitem[{\citenamefont{Bergmann et~al.}(1998)\citenamefont{Bergmann, Theuer,
  and Shore}}]{bergmann1998}
\bibinfo{author}{\bibfnamefont{K.}~\bibnamefont{Bergmann}},
  \bibinfo{author}{\bibfnamefont{H.}~\bibnamefont{Theuer}}, \bibnamefont{and}
  \bibinfo{author}{\bibfnamefont{B.~W.} \bibnamefont{Shore}},
  \bibinfo{journal}{Rev. Mod. Phys.} \textbf{\bibinfo{volume}{70}},
  \bibinfo{pages}{1003} (\bibinfo{year}{1998}).

\bibitem[{\citenamefont{Vitanov et~al.}(2001)\citenamefont{Vitanov,
  Fleischhauer, Shore, and Bergemann}}]{vitanov2001}
\bibinfo{author}{\bibfnamefont{N.~V.} \bibnamefont{Vitanov}},
  \bibinfo{author}{\bibfnamefont{M.}~\bibnamefont{Fleischhauer}},
  \bibinfo{author}{\bibfnamefont{B.~W.} \bibnamefont{Shore}}, \bibnamefont{and}
  \bibinfo{author}{\bibfnamefont{W.}~\bibnamefont{Bergemann}},
  \bibinfo{journal}{Adv. At. Mol. Opt. Phys.} \textbf{\bibinfo{volume}{46}},
  \bibinfo{pages}{55} (\bibinfo{year}{2001}).

\bibitem[{\citenamefont{Koch and Shapiro}(2012)}]{koch2012}
\bibinfo{author}{\bibfnamefont{C.~P.} \bibnamefont{Koch}} \bibnamefont{and}
  \bibinfo{author}{\bibfnamefont{M.}~\bibnamefont{Shapiro}},
  \bibinfo{journal}{Chem. Rev.} \textbf{\bibinfo{volume}{112}},
  \bibinfo{pages}{4928} (\bibinfo{year}{2012}).

\bibitem[{\citenamefont{Stuhl et~al.}(2012)\citenamefont{Stuhl, Hummon, Yeo,
  Qu\'em\'ener, Bohn, and Ye}}]{stuhl2012}
\bibinfo{author}{\bibfnamefont{B.~K.} \bibnamefont{Stuhl}},
  \bibinfo{author}{\bibfnamefont{M.~T.} \bibnamefont{Hummon}},
  \bibinfo{author}{\bibfnamefont{M.}~\bibnamefont{Yeo}},
  \bibinfo{author}{\bibfnamefont{G.}~\bibnamefont{Qu\'em\'ener}},
  \bibinfo{author}{\bibfnamefont{J.~L.} \bibnamefont{Bohn}}, \bibnamefont{and}
  \bibinfo{author}{\bibfnamefont{J.}~\bibnamefont{Ye}},
  \bibinfo{journal}{Nature} \textbf{\bibinfo{volume}{492}},
  \bibinfo{pages}{396} (\bibinfo{year}{2012}).

\bibitem[{\citenamefont{Egorov et~al.}(2004)\citenamefont{Egorov, Campbell,
  Friedrich, Maxwell, Tsikata, van Buuren, and Doyle}}]{egorov2004}
\bibinfo{author}{\bibfnamefont{D.}~\bibnamefont{Egorov}},
  \bibinfo{author}{\bibfnamefont{W.~C.} \bibnamefont{Campbell}},
  \bibinfo{author}{\bibfnamefont{B.}~\bibnamefont{Friedrich}},
  \bibinfo{author}{\bibfnamefont{S.~E.} \bibnamefont{Maxwell}},
  \bibinfo{author}{\bibfnamefont{E.}~\bibnamefont{Tsikata}},
  \bibinfo{author}{\bibfnamefont{L.~D.} \bibnamefont{van Buuren}},
  \bibnamefont{and} \bibinfo{author}{\bibfnamefont{J.~M.} \bibnamefont{Doyle}},
  \bibinfo{journal}{Eur. Phys. J. D} \textbf{\bibinfo{volume}{31}},
  \bibinfo{pages}{307} (\bibinfo{year}{2004}).

\bibitem[{\citenamefont{Hutzler et~al.}(2012)\citenamefont{Hutzler, Lu, and
  Doyle}}]{hutzler2012}
\bibinfo{author}{\bibfnamefont{N.~R.} \bibnamefont{Hutzler}},
  \bibinfo{author}{\bibfnamefont{H.-I.} \bibnamefont{Lu}}, \bibnamefont{and}
  \bibinfo{author}{\bibfnamefont{J.~M.} \bibnamefont{Doyle}},
  \bibinfo{journal}{Chem. Rev.} \textbf{\bibinfo{volume}{112}},
  \bibinfo{pages}{4803} (\bibinfo{year}{2012}).

\bibitem[{\citenamefont{Zeppenfeld et~al.}(2012)\citenamefont{Zeppenfeld,
  Englert, Gl\"ockner, Prehn, Mielenz, Sommer, van Buuren, Motsch, and
  Rempe}}]{zeppenfeld2012}
\bibinfo{author}{\bibfnamefont{M.}~\bibnamefont{Zeppenfeld}},
  \bibinfo{author}{\bibfnamefont{B.~G.~U.} \bibnamefont{Englert}},
  \bibinfo{author}{\bibfnamefont{R.}~\bibnamefont{Gl\"ockner}},
  \bibinfo{author}{\bibfnamefont{A.}~\bibnamefont{Prehn}},
  \bibinfo{author}{\bibfnamefont{M.}~\bibnamefont{Mielenz}},
  \bibinfo{author}{\bibfnamefont{C.}~\bibnamefont{Sommer}},
  \bibinfo{author}{\bibfnamefont{L.~D.} \bibnamefont{van Buuren}},
  \bibinfo{author}{\bibfnamefont{M.}~\bibnamefont{Motsch}}, \bibnamefont{and}
  \bibinfo{author}{\bibfnamefont{G.}~\bibnamefont{Rempe}},
  \bibinfo{journal}{Nature} \textbf{\bibinfo{volume}{491}},
  \bibinfo{pages}{570} (\bibinfo{year}{2012}).

\bibitem[{\citenamefont{Rosa}(2004)}]{dirosa2004}
\bibinfo{author}{\bibfnamefont{M.~D.~D.} \bibnamefont{Rosa}},
  \bibinfo{journal}{Eur. Phys. J. D} \textbf{\bibinfo{volume}{31}},
  \bibinfo{pages}{395} (\bibinfo{year}{2004}).

\bibitem[{\citenamefont{Hummon et~al.}(2013)\citenamefont{Hummon, Yeo, Stuhl,
  Collopy, Xia, and Ye}}]{hummon2013}
\bibinfo{author}{\bibfnamefont{M.~T.} \bibnamefont{Hummon}},
  \bibinfo{author}{\bibfnamefont{M.}~\bibnamefont{Yeo}},
  \bibinfo{author}{\bibfnamefont{B.~K.} \bibnamefont{Stuhl}},
  \bibinfo{author}{\bibfnamefont{A.~L.} \bibnamefont{Collopy}},
  \bibinfo{author}{\bibfnamefont{Y.}~\bibnamefont{Xia}}, \bibnamefont{and}
  \bibinfo{author}{\bibfnamefont{J.}~\bibnamefont{Ye}}, \bibinfo{journal}{Phys.
  Rev. Lett.} \textbf{\bibinfo{volume}{110}}, \bibinfo{pages}{143001}
  (\bibinfo{year}{2013}).

\bibitem[{\citenamefont{Zhelyazkova et~al.}(2014)\citenamefont{Zhelyazkova,
  Cournol, Wall, Matsushima, Hudson, Hinds, Tarbutt, and
  Sauer}}]{zhelyazkova2014}
\bibinfo{author}{\bibfnamefont{V.}~\bibnamefont{Zhelyazkova}},
  \bibinfo{author}{\bibfnamefont{A.}~\bibnamefont{Cournol}},
  \bibinfo{author}{\bibfnamefont{T.~E.} \bibnamefont{Wall}},
  \bibinfo{author}{\bibfnamefont{A.}~\bibnamefont{Matsushima}},
  \bibinfo{author}{\bibfnamefont{J.~J.} \bibnamefont{Hudson}},
  \bibinfo{author}{\bibfnamefont{E.~A.} \bibnamefont{Hinds}},
  \bibinfo{author}{\bibfnamefont{M.~R.} \bibnamefont{Tarbutt}},
  \bibnamefont{and} \bibinfo{author}{\bibfnamefont{B.~E.} \bibnamefont{Sauer}},
  \bibinfo{journal}{Phys. Rev. A} \textbf{\bibinfo{volume}{89}},
  \bibinfo{pages}{053416} (\bibinfo{year}{2014}).

\bibitem[{\citenamefont{Barry et~al.}(2014)\citenamefont{Barry, McCarron,
  Norrgard, Steinecker, and DeMille}}]{barry2014}
\bibinfo{author}{\bibfnamefont{J.~F.} \bibnamefont{Barry}},
  \bibinfo{author}{\bibfnamefont{D.~J.} \bibnamefont{McCarron}},
  \bibinfo{author}{\bibfnamefont{E.~B.} \bibnamefont{Norrgard}},
  \bibinfo{author}{\bibfnamefont{M.~H.} \bibnamefont{Steinecker}},
  \bibnamefont{and} \bibinfo{author}{\bibfnamefont{D.}~\bibnamefont{DeMille}},
  \bibinfo{journal}{Nature} \textbf{\bibinfo{volume}{512}},
  \bibinfo{pages}{286} (\bibinfo{year}{2014}).

\bibitem[{\citenamefont{Kobayashi J.; Aikawa K.;
  Oasa~K.}(2014)}]{kobayashi2014}
\bibinfo{author}{\bibfnamefont{I.~S.} \bibnamefont{Kobayashi J.; Aikawa K.;
  Oasa~K.}}, \bibinfo{journal}{Phys. Rev. A} \textbf{\bibinfo{volume}{89}},
  \bibinfo{pages}{021401} (\bibinfo{year}{2014}).

\bibitem[{\citenamefont{Doyle et~al.}(2004)\citenamefont{Doyle, Friedrich,
  Krems, and Masnou-Seeuws}}]{doyle2004}
\bibinfo{author}{\bibfnamefont{J.}~\bibnamefont{Doyle}},
  \bibinfo{author}{\bibfnamefont{B.}~\bibnamefont{Friedrich}},
  \bibinfo{author}{\bibfnamefont{R.}~\bibnamefont{Krems}}, \bibnamefont{and}
  \bibinfo{author}{\bibfnamefont{F.}~\bibnamefont{Masnou-Seeuws}},
  \bibinfo{journal}{Eur. Phys. J. D} \textbf{\bibinfo{volume}{31}},
  \bibinfo{pages}{149} (\bibinfo{year}{2004}).

\bibitem[{\citenamefont{Carr and Ye}(2009)}]{carr2009}
\bibinfo{author}{\bibfnamefont{L.~D.} \bibnamefont{Carr}} \bibnamefont{and}
  \bibinfo{author}{\bibfnamefont{J.}~\bibnamefont{Ye}}, \bibinfo{journal}{New
  J. Phys.} \textbf{\bibinfo{volume}{11}}, \bibinfo{pages}{055009}
  (\bibinfo{year}{2009}).

\bibitem[{\citenamefont{Dulieu and Gabbanini}(2009)}]{dulieu2009}
\bibinfo{author}{\bibfnamefont{O.}~\bibnamefont{Dulieu}} \bibnamefont{and}
  \bibinfo{author}{\bibfnamefont{C.}~\bibnamefont{Gabbanini}},
  \bibinfo{journal}{Rep. Prog. Phys.} \textbf{\bibinfo{volume}{72}},
  \bibinfo{pages}{086401} (\bibinfo{year}{2009}).

\bibitem[{\citenamefont{Jin and Ye}(2012)}]{jin2012}
\bibinfo{author}{\bibfnamefont{D.}~\bibnamefont{Jin}} \bibnamefont{and}
  \bibinfo{author}{\bibfnamefont{J.}~\bibnamefont{Ye}}, \bibinfo{journal}{Chem.
  Rev.} \textbf{\bibinfo{volume}{112}}, \bibinfo{pages}{4801}
  (\bibinfo{year}{2012}).

\bibitem[{\citenamefont{Stan et~al.}(2004)\citenamefont{Stan, Zwierlein,
  Schunck, Raupach, and Ketterle}}]{stan2004}
\bibinfo{author}{\bibfnamefont{C.~A.} \bibnamefont{Stan}},
  \bibinfo{author}{\bibfnamefont{M.~W.} \bibnamefont{Zwierlein}},
  \bibinfo{author}{\bibfnamefont{C.~H.} \bibnamefont{Schunck}},
  \bibinfo{author}{\bibfnamefont{S.~M.~F.} \bibnamefont{Raupach}},
  \bibnamefont{and} \bibinfo{author}{\bibfnamefont{W.}~\bibnamefont{Ketterle}},
  \bibinfo{journal}{Phys. Rev. Lett.} \textbf{\bibinfo{volume}{93}},
  \bibinfo{pages}{143001} (\bibinfo{year}{2004}).

\bibitem[{\citenamefont{Wille et~al.}(2008)\citenamefont{Wille, Spiegelhalder,
  Kerner, Naik, Trenkwalder, Hendl, Schreck, Grimm, Tiecke, Walraven
  et~al.}}]{wille2008}
\bibinfo{author}{\bibfnamefont{E.}~\bibnamefont{Wille}},
  \bibinfo{author}{\bibfnamefont{F.~M.} \bibnamefont{Spiegelhalder}},
  \bibinfo{author}{\bibfnamefont{G.}~\bibnamefont{Kerner}},
  \bibinfo{author}{\bibfnamefont{D.}~\bibnamefont{Naik}},
  \bibinfo{author}{\bibfnamefont{A.}~\bibnamefont{Trenkwalder}},
  \bibinfo{author}{\bibfnamefont{G.}~\bibnamefont{Hendl}},
  \bibinfo{author}{\bibfnamefont{F.}~\bibnamefont{Schreck}},
  \bibinfo{author}{\bibfnamefont{R.}~\bibnamefont{Grimm}},
  \bibinfo{author}{\bibfnamefont{T.~G.} \bibnamefont{Tiecke}},
  \bibinfo{author}{\bibfnamefont{J.~T.~M.} \bibnamefont{Walraven}},
  \bibnamefont{et~al.}, \bibinfo{journal}{Phys. Rev. Lett.}
  \textbf{\bibinfo{volume}{100}}, \bibinfo{pages}{053201}
  (\bibinfo{year}{2008}).

\bibitem[{\citenamefont{Deh et~al.}(2008)\citenamefont{Deh, Marzok, Zimmermann,
  and Courteille}}]{deh2008}
\bibinfo{author}{\bibfnamefont{B.}~\bibnamefont{Deh}},
  \bibinfo{author}{\bibfnamefont{C.}~\bibnamefont{Marzok}},
  \bibinfo{author}{\bibfnamefont{C.}~\bibnamefont{Zimmermann}},
  \bibnamefont{and} \bibinfo{author}{\bibfnamefont{P.~W.}
  \bibnamefont{Courteille}}, \bibinfo{journal}{Phys. Rev. A}
  \textbf{\bibinfo{volume}{77}}, \bibinfo{pages}{010701}
  (\bibinfo{year}{2008}).

\bibitem[{\citenamefont{Marzok et~al.}(2009)\citenamefont{Marzok, Deh,
  Zimmermann, Courteille, Tiemann, Vanne, and Saenz}}]{marzok2009}
\bibinfo{author}{\bibfnamefont{C.}~\bibnamefont{Marzok}},
  \bibinfo{author}{\bibfnamefont{B.}~\bibnamefont{Deh}},
  \bibinfo{author}{\bibfnamefont{C.}~\bibnamefont{Zimmermann}},
  \bibinfo{author}{\bibfnamefont{P.~W.} \bibnamefont{Courteille}},
  \bibinfo{author}{\bibfnamefont{E.}~\bibnamefont{Tiemann}},
  \bibinfo{author}{\bibfnamefont{Y.~V.} \bibnamefont{Vanne}}, \bibnamefont{and}
  \bibinfo{author}{\bibfnamefont{A.}~\bibnamefont{Saenz}},
  \bibinfo{journal}{Phys. Rev. A} \textbf{\bibinfo{volume}{79}},
  \bibinfo{pages}{012717} (\bibinfo{year}{2009}).

\bibitem[{\citenamefont{Repp et~al.}(2013)\citenamefont{Repp, Pires, Ulmanis,
  Heck, Kuhnle, and Weidem\"uller}}]{repp2013}
\bibinfo{author}{\bibfnamefont{M.}~\bibnamefont{Repp}},
  \bibinfo{author}{\bibfnamefont{R.}~\bibnamefont{Pires}},
  \bibinfo{author}{\bibfnamefont{J.}~\bibnamefont{Ulmanis}},
  \bibinfo{author}{\bibfnamefont{R.}~\bibnamefont{Heck}},
  \bibinfo{author}{\bibfnamefont{E.~D.} \bibnamefont{Kuhnle}},
  \bibnamefont{and}
  \bibinfo{author}{\bibfnamefont{M.}~\bibnamefont{Weidem\"uller}},
  \bibinfo{journal}{Phys. Rev. A} \textbf{\bibinfo{volume}{87}},
  \bibinfo{pages}{010701(R)} (\bibinfo{year}{2013}).

\bibitem[{\citenamefont{Cho et~al.}(2013)\citenamefont{Cho, McCarron,
  K\"oppinger, Jenkin, Butler, Julienne, Blackley, Le~Sueur, Hutson, and
  Cornish}}]{cho2013}
\bibinfo{author}{\bibfnamefont{H.-W.} \bibnamefont{Cho}},
  \bibinfo{author}{\bibfnamefont{D.~J.} \bibnamefont{McCarron}},
  \bibinfo{author}{\bibfnamefont{M.~P.} \bibnamefont{K\"oppinger}},
  \bibinfo{author}{\bibfnamefont{D.~L.} \bibnamefont{Jenkin}},
  \bibinfo{author}{\bibfnamefont{K.~L.} \bibnamefont{Butler}},
  \bibinfo{author}{\bibfnamefont{P.~S.} \bibnamefont{Julienne}},
  \bibinfo{author}{\bibfnamefont{C.~L.} \bibnamefont{Blackley}},
  \bibinfo{author}{\bibfnamefont{C.~R.} \bibnamefont{Le~Sueur}},
  \bibinfo{author}{\bibfnamefont{J.~M.} \bibnamefont{Hutson}},
  \bibnamefont{and} \bibinfo{author}{\bibfnamefont{S.~L.}
  \bibnamefont{Cornish}}, \bibinfo{journal}{Phys. Rev. A}
  \textbf{\bibinfo{volume}{87}}, \bibinfo{pages}{010703}
  (\bibinfo{year}{2013}).

\bibitem[{\citenamefont{Wu et~al.}(2012)\citenamefont{Wu, Park, Ahmadi, Will,
  and Zwierlein}}]{wu2012}
\bibinfo{author}{\bibfnamefont{C.-H.} \bibnamefont{Wu}},
  \bibinfo{author}{\bibfnamefont{J.~W.} \bibnamefont{Park}},
  \bibinfo{author}{\bibfnamefont{P.}~\bibnamefont{Ahmadi}},
  \bibinfo{author}{\bibfnamefont{S.}~\bibnamefont{Will}}, \bibnamefont{and}
  \bibinfo{author}{\bibfnamefont{M.~W.} \bibnamefont{Zwierlein}},
  \bibinfo{journal}{Phys. Rev. Lett.} \textbf{\bibinfo{volume}{109}},
  \bibinfo{pages}{085301} (\bibinfo{year}{2012}).

\bibitem[{\citenamefont{Wang et~al.}(2013)\citenamefont{Wang, Xiong, Li, Wang,
  and Tiemann}}]{wang2013}
\bibinfo{author}{\bibfnamefont{F.}~\bibnamefont{Wang}},
  \bibinfo{author}{\bibfnamefont{D.}~\bibnamefont{Xiong}},
  \bibinfo{author}{\bibfnamefont{X.}~\bibnamefont{Li}},
  \bibinfo{author}{\bibfnamefont{D.}~\bibnamefont{Wang}}, \bibnamefont{and}
  \bibinfo{author}{\bibfnamefont{E.}~\bibnamefont{Tiemann}},
  \bibinfo{journal}{Phys. Rev. A} \textbf{\bibinfo{volume}{87}},
  \bibinfo{pages}{050702} (\bibinfo{year}{2013}).

\bibitem[{\citenamefont{Inouye et~al.}(2004)\citenamefont{Inouye, Goldwin,
  Olsen, Ticknor, Bohn, and Jin}}]{inouye2004}
\bibinfo{author}{\bibfnamefont{S.}~\bibnamefont{Inouye}},
  \bibinfo{author}{\bibfnamefont{J.}~\bibnamefont{Goldwin}},
  \bibinfo{author}{\bibfnamefont{M.~L.} \bibnamefont{Olsen}},
  \bibinfo{author}{\bibfnamefont{C.}~\bibnamefont{Ticknor}},
  \bibinfo{author}{\bibfnamefont{J.~L.} \bibnamefont{Bohn}}, \bibnamefont{and}
  \bibinfo{author}{\bibfnamefont{D.~S.} \bibnamefont{Jin}},
  \bibinfo{journal}{Phys. Rev. Lett.} \textbf{\bibinfo{volume}{93}},
  \bibinfo{pages}{183201} (\bibinfo{year}{2004}).

\bibitem[{\citenamefont{Patel et~al.}(2014)\citenamefont{Patel, Blackley,
  Cornish, and Hutson}}]{patel2014}
\bibinfo{author}{\bibfnamefont{H.~J.} \bibnamefont{Patel}},
  \bibinfo{author}{\bibfnamefont{C.~L.} \bibnamefont{Blackley}},
  \bibinfo{author}{\bibfnamefont{S.~L.} \bibnamefont{Cornish}},
  \bibnamefont{and} \bibinfo{author}{\bibfnamefont{J.~M.}
  \bibnamefont{Hutson}}, \bibinfo{journal}{Phys. Rev. A}
  \textbf{\bibinfo{volume}{90}}, \bibinfo{pages}{032716}
  (\bibinfo{year}{2014}).

\bibitem[{\citenamefont{Pilch et~al.}(2009)\citenamefont{Pilch, Lange,
  Prantner, Kerner, Ferlaino, N\"agerl, and grimm R.}}]{pilch2009}
\bibinfo{author}{\bibfnamefont{K.}~\bibnamefont{Pilch}},
  \bibinfo{author}{\bibfnamefont{A.~D.} \bibnamefont{Lange}},
  \bibinfo{author}{\bibfnamefont{A.}~\bibnamefont{Prantner}},
  \bibinfo{author}{\bibfnamefont{G.}~\bibnamefont{Kerner}},
  \bibinfo{author}{\bibfnamefont{F.}~\bibnamefont{Ferlaino}},
  \bibinfo{author}{\bibfnamefont{H.-C.} \bibnamefont{N\"agerl}},
  \bibnamefont{and} \bibinfo{author}{\bibnamefont{grimm R.}},
  \bibinfo{journal}{Phys. Rev. A} \textbf{\bibinfo{volume}{79}},
  \bibinfo{pages}{042718} (\bibinfo{year}{2009}).

\bibitem[{\citenamefont{Aldegunde et~al.}(2008)\citenamefont{Aldegunde,
  Rivington, .{Z}uchowski, and Hutson}}]{aldegunde2008}
\bibinfo{author}{\bibfnamefont{J.}~\bibnamefont{Aldegunde}},
  \bibinfo{author}{\bibfnamefont{B.~A.} \bibnamefont{Rivington}},
  \bibinfo{author}{\bibfnamefont{P.~S.} \bibnamefont{.{Z}uchowski}},
  \bibnamefont{and} \bibinfo{author}{\bibfnamefont{J.~M.}
  \bibnamefont{Hutson}}, \bibinfo{journal}{Physical Review A}
  \textbf{\bibinfo{volume}{78}}, \bibinfo{pages}{033434}
  (\bibinfo{year}{2008}).

\bibitem[{\citenamefont{Aldegunde and Hutson}(2009)}]{aldegunde2009}
\bibinfo{author}{\bibfnamefont{J.}~\bibnamefont{Aldegunde}} \bibnamefont{and}
  \bibinfo{author}{\bibfnamefont{J.~M.} \bibnamefont{Hutson}},
  \bibinfo{journal}{Physical Review A} \textbf{\bibinfo{volume}{79}},
  \bibinfo{pages}{013401} (\bibinfo{year}{2009}).

\bibitem[{\citenamefont{Ran et~al.}(2010)\citenamefont{Ran, Aldegunde, and
  Hutson}}]{ran2010}
\bibinfo{author}{\bibfnamefont{H.}~\bibnamefont{Ran}},
  \bibinfo{author}{\bibfnamefont{J.}~\bibnamefont{Aldegunde}},
  \bibnamefont{and} \bibinfo{author}{\bibfnamefont{J.~M.}
  \bibnamefont{Hutson}}, \bibinfo{journal}{New J. Phys.}
  \textbf{\bibinfo{volume}{12}}, \bibinfo{pages}{043015}
  (\bibinfo{year}{2010}).

\bibitem[{\citenamefont{Ospelkaus et~al.}(2010)\citenamefont{Ospelkaus, Ni,
  Qu\'em\'ener, Neyenhuis, Wang, deMiranda, Bohn, Ye, and
  Jin}}]{ospelkaus2010b}
\bibinfo{author}{\bibfnamefont{S.}~\bibnamefont{Ospelkaus}},
  \bibinfo{author}{\bibfnamefont{K.-K.} \bibnamefont{Ni}},
  \bibinfo{author}{\bibfnamefont{G.}~\bibnamefont{Qu\'em\'ener}},
  \bibinfo{author}{\bibfnamefont{B.}~\bibnamefont{Neyenhuis}},
  \bibinfo{author}{\bibfnamefont{D.}~\bibnamefont{Wang}},
  \bibinfo{author}{\bibfnamefont{M.~H.~G.} \bibnamefont{deMiranda}},
  \bibinfo{author}{\bibfnamefont{J.~L.} \bibnamefont{Bohn}},
  \bibinfo{author}{\bibfnamefont{J.}~\bibnamefont{Ye}}, \bibnamefont{and}
  \bibinfo{author}{\bibfnamefont{D.}~\bibnamefont{Jin}},
  \bibinfo{journal}{Phys. Rev. Lett.} \textbf{\bibinfo{volume}{104}},
  \bibinfo{pages}{030402} (\bibinfo{year}{2010}).

\bibitem[{\citenamefont{Strauss et~al.}(2010)\citenamefont{Strauss, Takekoshi,
  Lang, Winkler, Grimm, Denschlag, and Tieman}}]{strauss2010}
\bibinfo{author}{\bibfnamefont{C.}~\bibnamefont{Strauss}},
  \bibinfo{author}{\bibfnamefont{T.}~\bibnamefont{Takekoshi}},
  \bibinfo{author}{\bibfnamefont{F.}~\bibnamefont{Lang}},
  \bibinfo{author}{\bibfnamefont{K.}~\bibnamefont{Winkler}},
  \bibinfo{author}{\bibfnamefont{R.}~\bibnamefont{Grimm}},
  \bibinfo{author}{\bibfnamefont{J.~H.} \bibnamefont{Denschlag}},
  \bibnamefont{and} \bibinfo{author}{\bibfnamefont{E.}~\bibnamefont{Tieman}},
  \bibinfo{journal}{Phys. Rev. A} \textbf{\bibinfo{volume}{82}},
  \bibinfo{pages}{052514} (\bibinfo{year}{2010}).

\bibitem[{\citenamefont{Wall et~al.}(2013)\citenamefont{Wall, Bekaroglu, and
  Carr}}]{wall2013}
\bibinfo{author}{\bibfnamefont{M.~L.} \bibnamefont{Wall}},
  \bibinfo{author}{\bibfnamefont{E.}~\bibnamefont{Bekaroglu}},
  \bibnamefont{and} \bibinfo{author}{\bibfnamefont{L.~D.} \bibnamefont{Carr}},
  \bibinfo{journal}{Phys. Rev. A} \textbf{\bibinfo{volume}{88}},
  \bibinfo{pages}{023605} (\bibinfo{year}{2013}).

\bibitem[{\citenamefont{Stwalley et~al.}(1978)\citenamefont{Stwalley, Uang, and
  Pichler}}]{stwalley1978}
\bibinfo{author}{\bibfnamefont{W.~C.} \bibnamefont{Stwalley}},
  \bibinfo{author}{\bibfnamefont{Y.~H.} \bibnamefont{Uang}}, \bibnamefont{and}
  \bibinfo{author}{\bibfnamefont{G.}~\bibnamefont{Pichler}},
  \bibinfo{journal}{Phys. Rev. Lett.} \textbf{\bibinfo{volume}{41}},
  \bibinfo{pages}{1164} (\bibinfo{year}{1978}).

\bibitem[{\citenamefont{Movre and Pichler}(1977)}]{movre1977}
\bibinfo{author}{\bibfnamefont{M.}~\bibnamefont{Movre}} \bibnamefont{and}
  \bibinfo{author}{\bibfnamefont{G.}~\bibnamefont{Pichler}},
  \bibinfo{journal}{J. Phys. B: At. Mol. Opt. Phys.}
  \textbf{\bibinfo{volume}{10}}, \bibinfo{pages}{2631} (\bibinfo{year}{1977}).

\bibitem[{\citenamefont{Comparat et~al.}(2000)\citenamefont{Comparat, Drag,
  Tolra, Fioretti, Pillet, Crubellier, Dulieu, and
  Masnou-Seeuws}}]{comparat2000}
\bibinfo{author}{\bibfnamefont{D.}~\bibnamefont{Comparat}},
  \bibinfo{author}{\bibfnamefont{C.}~\bibnamefont{Drag}},
  \bibinfo{author}{\bibfnamefont{B.~L.} \bibnamefont{Tolra}},
  \bibinfo{author}{\bibfnamefont{A.}~\bibnamefont{Fioretti}},
  \bibinfo{author}{\bibfnamefont{P.}~\bibnamefont{Pillet}},
  \bibinfo{author}{\bibfnamefont{A.}~\bibnamefont{Crubellier}},
  \bibinfo{author}{\bibfnamefont{O.}~\bibnamefont{Dulieu}}, \bibnamefont{and}
  \bibinfo{author}{\bibfnamefont{F.}~\bibnamefont{Masnou-Seeuws}},
  \bibinfo{journal}{Eur. Phys. J. D} \textbf{\bibinfo{volume}{11}},
  \bibinfo{pages}{59} (\bibinfo{year}{2000}).

\bibitem[{\citenamefont{Kemmann et~al.}(2004)\citenamefont{Kemmann, Mistrik,
  Nussmann, Helm, Williams, and Julienne}}]{kemmann2004}
\bibinfo{author}{\bibfnamefont{M.}~\bibnamefont{Kemmann}},
  \bibinfo{author}{\bibfnamefont{I.}~\bibnamefont{Mistrik}},
  \bibinfo{author}{\bibfnamefont{S.}~\bibnamefont{Nussmann}},
  \bibinfo{author}{\bibfnamefont{H.}~\bibnamefont{Helm}},
  \bibinfo{author}{\bibfnamefont{C.~J.} \bibnamefont{Williams}},
  \bibnamefont{and} \bibinfo{author}{\bibfnamefont{P.~S.}
  \bibnamefont{Julienne}}, \bibinfo{journal}{Phys. Rev. A}
  \textbf{\bibinfo{volume}{69}}, \bibinfo{pages}{022715}
  (\bibinfo{year}{2004}).

\bibitem[{\citenamefont{Tiesinga et~al.}(2005)\citenamefont{Tiesinga, Jones,
  Lett, Volz, Williams, and Julienne}}]{tiesinga2005}
\bibinfo{author}{\bibfnamefont{E.}~\bibnamefont{Tiesinga}},
  \bibinfo{author}{\bibfnamefont{K.~M.} \bibnamefont{Jones}},
  \bibinfo{author}{\bibfnamefont{P.~D.} \bibnamefont{Lett}},
  \bibinfo{author}{\bibfnamefont{U.}~\bibnamefont{Volz}},
  \bibinfo{author}{\bibfnamefont{C.~J.} \bibnamefont{Williams}},
  \bibnamefont{and} \bibinfo{author}{\bibfnamefont{P.~S.}
  \bibnamefont{Julienne}}, \bibinfo{journal}{Phys. Rev. A}
  \textbf{\bibinfo{volume}{71}}, \bibinfo{pages}{052703}
  (\bibinfo{year}{2005}).

\bibitem[{\citenamefont{Kasahara et~al.}(1996)\citenamefont{Kasahara, Ebi,
  M.Tanimura, Ikoma, Matsubara, Baba, and Kat\^{o}}}]{kasahara1996}
\bibinfo{author}{\bibfnamefont{S.}~\bibnamefont{Kasahara}},
  \bibinfo{author}{\bibfnamefont{T.}~\bibnamefont{Ebi}},
  \bibinfo{author}{\bibnamefont{M.Tanimura}},
  \bibinfo{author}{\bibfnamefont{H.}~\bibnamefont{Ikoma}},
  \bibinfo{author}{\bibfnamefont{K.}~\bibnamefont{Matsubara}},
  \bibinfo{author}{\bibfnamefont{M.}~\bibnamefont{Baba}}, \bibnamefont{and}
  \bibinfo{author}{\bibfnamefont{H.}~\bibnamefont{Kat\^{o}}},
  \bibinfo{journal}{J. Chem. Phys.} \textbf{\bibinfo{volume}{105}},
  \bibinfo{pages}{1341} (\bibinfo{year}{1996}).

\bibitem[{\citenamefont{Burns et~al.}(2003)\citenamefont{Burns, Sibbach-Morgus,
  Wilkins, Halpern, Clarke, Miles, Li, Hickman, and Huennekens}}]{burns2003}
\bibinfo{author}{\bibfnamefont{P.}~\bibnamefont{Burns}},
  \bibinfo{author}{\bibfnamefont{L.}~\bibnamefont{Sibbach-Morgus}},
  \bibinfo{author}{\bibfnamefont{A.~D.} \bibnamefont{Wilkins}},
  \bibinfo{author}{\bibfnamefont{F.}~\bibnamefont{Halpern}},
  \bibinfo{author}{\bibfnamefont{L.}~\bibnamefont{Clarke}},
  \bibinfo{author}{\bibfnamefont{R.~D.} \bibnamefont{Miles}},
  \bibinfo{author}{\bibfnamefont{L.}~\bibnamefont{Li}},
  \bibinfo{author}{\bibfnamefont{A.~P.} \bibnamefont{Hickman}},
  \bibnamefont{and}
  \bibinfo{author}{\bibfnamefont{J.}~\bibnamefont{Huennekens}},
  \bibinfo{journal}{The Journal of Chemical Physics}
  \textbf{\bibinfo{volume}{119}}, \bibinfo{pages}{4743} (\bibinfo{year}{2003}).

\bibitem[{\citenamefont{Burns et~al.}(2005)\citenamefont{Burns, Wilkins,
  Hickman, and Huennekens}}]{burns2005}
\bibinfo{author}{\bibfnamefont{P.}~\bibnamefont{Burns}},
  \bibinfo{author}{\bibfnamefont{A.~D.} \bibnamefont{Wilkins}},
  \bibinfo{author}{\bibfnamefont{A.~P.} \bibnamefont{Hickman}},
  \bibnamefont{and}
  \bibinfo{author}{\bibfnamefont{J.}~\bibnamefont{Huennekens}},
  \bibinfo{journal}{The Journal of Chemical Physics}
  \textbf{\bibinfo{volume}{122}}, \bibinfo{pages}{074306}
  (\bibinfo{year}{2005}).

\bibitem[{\citenamefont{Wilkins et~al.}(2005)\citenamefont{Wilkins, Morgus,
  Hernandez-Guzman, Huennekens, and Hickman}}]{wilkins2005}
\bibinfo{author}{\bibfnamefont{A.~D.} \bibnamefont{Wilkins}},
  \bibinfo{author}{\bibfnamefont{L.}~\bibnamefont{Morgus}},
  \bibinfo{author}{\bibfnamefont{J.}~\bibnamefont{Hernandez-Guzman}},
  \bibinfo{author}{\bibfnamefont{J.}~\bibnamefont{Huennekens}},
  \bibnamefont{and} \bibinfo{author}{\bibfnamefont{A.~P.}
  \bibnamefont{Hickman}}, \bibinfo{journal}{J. Chem. Phys.}
  \textbf{\bibinfo{volume}{123}}, \bibinfo{pages}{124306}
  (\bibinfo{year}{2005}).

\bibitem[{\citenamefont{Takekoshi et~al.}(2011)\citenamefont{Takekoshi,
  Strauss, Lang, Denschlag, Lysebo, and Veseth}}]{takekoshi2011}
\bibinfo{author}{\bibfnamefont{T.}~\bibnamefont{Takekoshi}},
  \bibinfo{author}{\bibfnamefont{C.}~\bibnamefont{Strauss}},
  \bibinfo{author}{\bibfnamefont{F.}~\bibnamefont{Lang}},
  \bibinfo{author}{\bibfnamefont{J.~H.} \bibnamefont{Denschlag}},
  \bibinfo{author}{\bibfnamefont{M.}~\bibnamefont{Lysebo}}, \bibnamefont{and}
  \bibinfo{author}{\bibfnamefont{L.}~\bibnamefont{Veseth}},
  \bibinfo{journal}{Phys. Rev. A} \textbf{\bibinfo{volume}{83}},
  \bibinfo{pages}{062504} (\bibinfo{year}{2011}).

\bibitem[{\citenamefont{Debatin et~al.}(2011)\citenamefont{Debatin, Takekoshi,
  Rameshan, Reichsoellner, Ferlaino, Grimm, Vexiau, Bouloufa, Dulieu, and
  N\"agerl}}]{debatin2011}
\bibinfo{author}{\bibfnamefont{M.}~\bibnamefont{Debatin}},
  \bibinfo{author}{\bibfnamefont{T.}~\bibnamefont{Takekoshi}},
  \bibinfo{author}{\bibfnamefont{R.}~\bibnamefont{Rameshan}},
  \bibinfo{author}{\bibfnamefont{L.}~\bibnamefont{Reichsoellner}},
  \bibinfo{author}{\bibfnamefont{F.}~\bibnamefont{Ferlaino}},
  \bibinfo{author}{\bibfnamefont{R.}~\bibnamefont{Grimm}},
  \bibinfo{author}{\bibfnamefont{R.}~\bibnamefont{Vexiau}},
  \bibinfo{author}{\bibfnamefont{N.}~\bibnamefont{Bouloufa}},
  \bibinfo{author}{\bibfnamefont{O.}~\bibnamefont{Dulieu}}, \bibnamefont{and}
  \bibinfo{author}{\bibfnamefont{H.-C.} \bibnamefont{N\"agerl}},
  \bibinfo{journal}{Phys. Chem. Chem. Phys.} \textbf{\bibinfo{volume}{13}},
  \bibinfo{pages}{18926} (\bibinfo{year}{2011}).

\bibitem[{\citenamefont{Lysebo and Veseth}(2013)}]{lysebo2013}
\bibinfo{author}{\bibfnamefont{M.}~\bibnamefont{Lysebo}} \bibnamefont{and}
  \bibinfo{author}{\bibfnamefont{L.}~\bibnamefont{Veseth}},
  \bibinfo{journal}{Eur. Phys. J. D} \textbf{\bibinfo{volume}{67}},
  \bibinfo{pages}{142} (\bibinfo{year}{2013}).

\bibitem[{\citenamefont{\ifmmode~\dot{Z}\else \.{Z}\fi{}uchowski
  et~al.}(2010)\citenamefont{\ifmmode~\dot{Z}\else \.{Z}\fi{}uchowski,
  Aldegunde, and Hutson}}]{zuchowski2010a}
\bibinfo{author}{\bibfnamefont{P.~S.} \bibnamefont{\ifmmode~\dot{Z}\else
  \.{Z}\fi{}uchowski}},
  \bibinfo{author}{\bibfnamefont{J.}~\bibnamefont{Aldegunde}},
  \bibnamefont{and} \bibinfo{author}{\bibfnamefont{J.~M.}
  \bibnamefont{Hutson}}, \bibinfo{journal}{Phys. Rev. Lett.}
  \textbf{\bibinfo{volume}{105}}, \bibinfo{pages}{153201}
  (\bibinfo{year}{2010}).

\bibitem[{\citenamefont{Broyer et~al.}(1978)\citenamefont{Broyer, Vigu\'e, and
  Lehmann}}]{broyer1978}
\bibinfo{author}{\bibfnamefont{M.}~\bibnamefont{Broyer}},
  \bibinfo{author}{\bibfnamefont{J.}~\bibnamefont{Vigu\'e}}, \bibnamefont{and}
  \bibinfo{author}{\bibfnamefont{J.}~\bibnamefont{Lehmann}},
  \bibinfo{journal}{J. Phys.} \textbf{\bibinfo{volume}{39}},
  \bibinfo{pages}{591} (\bibinfo{year}{1978}).

\bibitem[{\citenamefont{Kruzins et~al.}(2010)\citenamefont{Kruzins, Klincare,
  Nikolayeva, Tamanis, Ferber, Pazyuk, and Stolyarov}}]{kruzins2010}
\bibinfo{author}{\bibfnamefont{A.}~\bibnamefont{Kruzins}},
  \bibinfo{author}{\bibfnamefont{I.}~\bibnamefont{Klincare}},
  \bibinfo{author}{\bibfnamefont{O.}~\bibnamefont{Nikolayeva}},
  \bibinfo{author}{\bibfnamefont{M.}~\bibnamefont{Tamanis}},
  \bibinfo{author}{\bibfnamefont{R.}~\bibnamefont{Ferber}},
  \bibinfo{author}{\bibfnamefont{E.~A.} \bibnamefont{Pazyuk}},
  \bibnamefont{and} \bibinfo{author}{\bibfnamefont{A.~V.}
  \bibnamefont{Stolyarov}}, \bibinfo{journal}{Phys. Rev. A}
  \textbf{\bibinfo{volume}{81}}, \bibinfo{pages}{042509}
  (\bibinfo{year}{2010}).

\bibitem[{\citenamefont{Kruzins et~al.}(2013)\citenamefont{Kruzins, Klincare,
  Nikolayeva, Tamanis, Ferber, Pazyuk, and Stolyarov}}]{kruzins2013}
\bibinfo{author}{\bibfnamefont{A.}~\bibnamefont{Kruzins}},
  \bibinfo{author}{\bibfnamefont{I.}~\bibnamefont{Klincare}},
  \bibinfo{author}{\bibfnamefont{O.}~\bibnamefont{Nikolayeva}},
  \bibinfo{author}{\bibfnamefont{M.}~\bibnamefont{Tamanis}},
  \bibinfo{author}{\bibfnamefont{R.}~\bibnamefont{Ferber}},
  \bibinfo{author}{\bibfnamefont{E.~A.} \bibnamefont{Pazyuk}},
  \bibnamefont{and} \bibinfo{author}{\bibfnamefont{A.~V.}
  \bibnamefont{Stolyarov}}, \bibinfo{journal}{J. Chem. Phys.}
  \textbf{\bibinfo{volume}{139}}, \bibinfo{pages}{244301}
  (\bibinfo{year}{2013}).

\bibitem[{\citenamefont{Aymar and Dulieu}(2005)}]{aymar2005}
\bibinfo{author}{\bibfnamefont{M.}~\bibnamefont{Aymar}} \bibnamefont{and}
  \bibinfo{author}{\bibfnamefont{O.}~\bibnamefont{Dulieu}},
  \bibinfo{journal}{J. Chem. Phys.} \textbf{\bibinfo{volume}{122}},
  \bibinfo{pages}{204302} (\bibinfo{year}{2005}).

\bibitem[{\citenamefont{Aymar and Dulieu}(2006)}]{aymar2006a}
\bibinfo{author}{\bibfnamefont{M.}~\bibnamefont{Aymar}} \bibnamefont{and}
  \bibinfo{author}{\bibfnamefont{O.}~\bibnamefont{Dulieu}},
  \bibinfo{journal}{J. Chem. Phys.} \textbf{\bibinfo{volume}{125}},
  \bibinfo{pages}{047101} (\bibinfo{year}{2006}).

\bibitem[{\citenamefont{Gu\'erout et~al.}(2010)\citenamefont{Gu\'erout, Aymar,
  and Dulieu}}]{guerout2010}
\bibinfo{author}{\bibfnamefont{R.}~\bibnamefont{Gu\'erout}},
  \bibinfo{author}{\bibfnamefont{M.}~\bibnamefont{Aymar}}, \bibnamefont{and}
  \bibinfo{author}{\bibfnamefont{O.}~\bibnamefont{Dulieu}},
  \bibinfo{journal}{Phys. Rev. A} \textbf{\bibinfo{volume}{82}},
  \bibinfo{pages}{042508} (\bibinfo{year}{2010}).

\bibitem[{\citenamefont{Borsalino et~al.}(2015)\citenamefont{Borsalino, Vexiau,
  Aymar, Luc-Koenig, Dulieu, and Bouloufa-Maafa}}]{borsalino2015}
\bibinfo{author}{\bibfnamefont{D.}~\bibnamefont{Borsalino}},
  \bibinfo{author}{\bibfnamefont{R.}~\bibnamefont{Vexiau}},
  \bibinfo{author}{\bibfnamefont{M.}~\bibnamefont{Aymar}},
  \bibinfo{author}{\bibfnamefont{E.}~\bibnamefont{Luc-Koenig}},
  \bibinfo{author}{\bibfnamefont{O.}~\bibnamefont{Dulieu}}, \bibnamefont{and}
  \bibinfo{author}{\bibfnamefont{N.}~\bibnamefont{Bouloufa-Maafa}},
  \bibinfo{journal}{arXiv:1501.06276}  (\bibinfo{year}{2015}).

\bibitem[{\citenamefont{Ralchenko et~al.}(2014)\citenamefont{Ralchenko,
  Kramida, J.~Reader, and Team}}]{nist_database}
\bibinfo{author}{\bibfnamefont{Y.}~\bibnamefont{Ralchenko}},
  \bibinfo{author}{\bibfnamefont{A.}~\bibnamefont{Kramida}},
  \bibinfo{author}{\bibfnamefont{J.}~\bibnamefont{J.~Reader}},
  \bibnamefont{and} \bibinfo{author}{\bibfnamefont{N.~A.} \bibnamefont{Team}}
  (\bibinfo{publisher}{National Institute of Standards and Technology},
  \bibinfo{address}{Gaithersburg, MD.}, \bibinfo{year}{2014}).

\bibitem[{\citenamefont{Tamanis et~al.}(2010)\citenamefont{Tamanis, Klincare,
  Kruzins, Nikolayeva, Ferber, Pazyuk, and Stolyarov}}]{tamanis2010}
\bibinfo{author}{\bibfnamefont{M.}~\bibnamefont{Tamanis}},
  \bibinfo{author}{\bibfnamefont{I.}~\bibnamefont{Klincare}},
  \bibinfo{author}{\bibfnamefont{A.}~\bibnamefont{Kruzins}},
  \bibinfo{author}{\bibfnamefont{O.}~\bibnamefont{Nikolayeva}},
  \bibinfo{author}{\bibfnamefont{R.}~\bibnamefont{Ferber}},
  \bibinfo{author}{\bibfnamefont{E.~A.} \bibnamefont{Pazyuk}},
  \bibnamefont{and} \bibinfo{author}{\bibfnamefont{A.~V.}
  \bibnamefont{Stolyarov}}, \bibinfo{journal}{Phys. Rev. A}
  \textbf{\bibinfo{volume}{82}}, \bibinfo{pages}{032506}
  (\bibinfo{year}{2010}).

\bibitem[{\citenamefont{Kim et~al.}(2009)\citenamefont{Kim, Lee, and
  Stolyarov}}]{kim2009}
\bibinfo{author}{\bibfnamefont{J.}~\bibnamefont{Kim}},
  \bibinfo{author}{\bibfnamefont{Y.}~\bibnamefont{Lee}}, \bibnamefont{and}
  \bibinfo{author}{\bibfnamefont{A.}~\bibnamefont{Stolyarov}},
  \bibinfo{journal}{J. Mol. Spectrosc.} \textbf{\bibinfo{volume}{256}},
  \bibinfo{pages}{57} (\bibinfo{year}{2009}).

\bibitem[{\citenamefont{Weiner et~al.}(1999)\citenamefont{Weiner, Bagnato,
  Zilio, and Julienne}}]{weiner1999}
\bibinfo{author}{\bibfnamefont{J.}~\bibnamefont{Weiner}},
  \bibinfo{author}{\bibfnamefont{V.~S.} \bibnamefont{Bagnato}},
  \bibinfo{author}{\bibfnamefont{S.~C.} \bibnamefont{Zilio}}, \bibnamefont{and}
  \bibinfo{author}{\bibfnamefont{P.~S.} \bibnamefont{Julienne}},
  \bibinfo{journal}{Rev. Mod. Phys.} \textbf{\bibinfo{volume}{71}},
  \bibinfo{pages}{1} (\bibinfo{year}{1999}).

\bibitem[{\citenamefont{Bali et~al.}(1994)\citenamefont{Bali, Hoffmann, and
  Walker}}]{bali1994}
\bibinfo{author}{\bibfnamefont{S.}~\bibnamefont{Bali}},
  \bibinfo{author}{\bibfnamefont{D.}~\bibnamefont{Hoffmann}}, \bibnamefont{and}
  \bibinfo{author}{\bibfnamefont{T.}~\bibnamefont{Walker}},
  \bibinfo{journal}{Europhys. Lett.} \textbf{\bibinfo{volume}{27}},
  \bibinfo{pages}{273} (\bibinfo{year}{1994}).

\bibitem[{\citenamefont{Sanchez-Villicana
  et~al.}(1995)\citenamefont{Sanchez-Villicana, Gensemer, Tan, Kumarakrishnan,
  Dinneen, S\"uptitz, and Gould}}]{sanchez-villicana1995}
\bibinfo{author}{\bibfnamefont{V.}~\bibnamefont{Sanchez-Villicana}},
  \bibinfo{author}{\bibfnamefont{S.~D.} \bibnamefont{Gensemer}},
  \bibinfo{author}{\bibfnamefont{K.~Y.~N.} \bibnamefont{Tan}},
  \bibinfo{author}{\bibfnamefont{A.}~\bibnamefont{Kumarakrishnan}},
  \bibinfo{author}{\bibfnamefont{T.~P.} \bibnamefont{Dinneen}},
  \bibinfo{author}{\bibfnamefont{W.}~\bibnamefont{S\"uptitz}},
  \bibnamefont{and} \bibinfo{author}{\bibfnamefont{P.~L.} \bibnamefont{Gould}},
  \bibinfo{journal}{Phys. Rev. Lett.} \textbf{\bibinfo{volume}{74}},
  \bibinfo{pages}{4619} (\bibinfo{year}{1995}).

\bibitem[{\citenamefont{Zilio et~al.}(1996)\citenamefont{Zilio, Marcassa,
  Muniz, Horowicz, Bagnato, Napolitano, Weiner, and Julienne}}]{zilio1996}
\bibinfo{author}{\bibfnamefont{S.}~\bibnamefont{Zilio}},
  \bibinfo{author}{\bibfnamefont{L.}~\bibnamefont{Marcassa}},
  \bibinfo{author}{\bibfnamefont{S.}~\bibnamefont{Muniz}},
  \bibinfo{author}{\bibfnamefont{R.}~\bibnamefont{Horowicz}},
  \bibinfo{author}{\bibfnamefont{V.}~\bibnamefont{Bagnato}},
  \bibinfo{author}{\bibfnamefont{R.}~\bibnamefont{Napolitano}},
  \bibinfo{author}{\bibfnamefont{J.}~\bibnamefont{Weiner}}, \bibnamefont{and}
  \bibinfo{author}{\bibfnamefont{P.~S.} \bibnamefont{Julienne}},
  \bibinfo{journal}{Phys. Rev. Lett.} \textbf{\bibinfo{volume}{76}},
  \bibinfo{pages}{2033} (\bibinfo{year}{1996}).

\bibitem[{\citenamefont{Sukenik et~al.}(1998)\citenamefont{Sukenik, Hoffmann,
  Bali, and Walker}}]{sukenik1998}
\bibinfo{author}{\bibfnamefont{C.~I.} \bibnamefont{Sukenik}},
  \bibinfo{author}{\bibfnamefont{D.}~\bibnamefont{Hoffmann}},
  \bibinfo{author}{\bibfnamefont{S.}~\bibnamefont{Bali}}, \bibnamefont{and}
  \bibinfo{author}{\bibfnamefont{T.}~\bibnamefont{Walker}},
  \bibinfo{journal}{Phys. Rev. Lett.} \textbf{\bibinfo{volume}{81}},
  \bibinfo{pages}{782} (\bibinfo{year}{1998}).

\bibitem[{\citenamefont{Walhout et~al.}(1995)\citenamefont{Walhout, Sterr,
  Orzel, Hoogerland, and Rolston}}]{walhout1995}
\bibinfo{author}{\bibfnamefont{M.}~\bibnamefont{Walhout}},
  \bibinfo{author}{\bibfnamefont{U.}~\bibnamefont{Sterr}},
  \bibinfo{author}{\bibfnamefont{C.}~\bibnamefont{Orzel}},
  \bibinfo{author}{\bibfnamefont{M.}~\bibnamefont{Hoogerland}},
  \bibnamefont{and} \bibinfo{author}{\bibfnamefont{S.}~\bibnamefont{Rolston}},
  \bibinfo{journal}{Phys. Rev. Lett.} \textbf{\bibinfo{volume}{74}},
  \bibinfo{pages}{506} (\bibinfo{year}{1995}).

\bibitem[{\citenamefont{Katori and Shimizu}(1994)}]{katori1994}
\bibinfo{author}{\bibfnamefont{H.}~\bibnamefont{Katori}} \bibnamefont{and}
  \bibinfo{author}{\bibfnamefont{F.}~\bibnamefont{Shimizu}},
  \bibinfo{journal}{Phys. Rev. Lett.} \textbf{\bibinfo{volume}{73}},
  \bibinfo{pages}{2555} (\bibinfo{year}{1994}).

\bibitem[{\citenamefont{Marcassa et~al.}(1994)\citenamefont{Marcassa, Muniz,
  de~Queiroz, Zilio, Bagnato, Weiner, Julienne, and Suominen}}]{marcassa1994}
\bibinfo{author}{\bibfnamefont{L.}~\bibnamefont{Marcassa}},
  \bibinfo{author}{\bibfnamefont{S.}~\bibnamefont{Muniz}},
  \bibinfo{author}{\bibfnamefont{E.}~\bibnamefont{de~Queiroz}},
  \bibinfo{author}{\bibfnamefont{S.}~\bibnamefont{Zilio}},
  \bibinfo{author}{\bibfnamefont{V.}~\bibnamefont{Bagnato}},
  \bibinfo{author}{\bibfnamefont{J.}~\bibnamefont{Weiner}},
  \bibinfo{author}{\bibfnamefont{P.~S.} \bibnamefont{Julienne}},
  \bibnamefont{and} \bibinfo{author}{\bibfnamefont{K.~A.}
  \bibnamefont{Suominen}}, \bibinfo{journal}{Phys. Rev. Lett.}
  \textbf{\bibinfo{volume}{73}}, \bibinfo{pages}{1911} (\bibinfo{year}{1994}).

\bibitem[{\citenamefont{Zirbel et~al.}(2008)\citenamefont{Zirbel, Ni,
  Ospelkaus, Nicholson, Olsen, Julienne, Wieman, Ye, and Jin}}]{zirbel2008}
\bibinfo{author}{\bibfnamefont{J.~J.} \bibnamefont{Zirbel}},
  \bibinfo{author}{\bibfnamefont{K.-K.} \bibnamefont{Ni}},
  \bibinfo{author}{\bibfnamefont{S.}~\bibnamefont{Ospelkaus}},
  \bibinfo{author}{\bibfnamefont{T.~L.} \bibnamefont{Nicholson}},
  \bibinfo{author}{\bibfnamefont{M.~L.} \bibnamefont{Olsen}},
  \bibinfo{author}{\bibfnamefont{P.~S.} \bibnamefont{Julienne}},
  \bibinfo{author}{\bibfnamefont{C.~E.} \bibnamefont{Wieman}},
  \bibinfo{author}{\bibfnamefont{J.}~\bibnamefont{Ye}}, \bibnamefont{and}
  \bibinfo{author}{\bibfnamefont{D.~S.} \bibnamefont{Jin}},
  \bibinfo{journal}{Phys. Rev. A} \textbf{\bibinfo{volume}{78}},
  \bibinfo{pages}{013416} (\bibinfo{year}{2008}).

\bibitem[{\citenamefont{Borsalino et~al.}(2014)\citenamefont{Borsalino,
  Londo\~no Flor\`ez, Vexiau, Dulieu, Bouloufa-Maafa, and
  Luc-Koenig}}]{borsalino2014}
\bibinfo{author}{\bibfnamefont{D.}~\bibnamefont{Borsalino}},
  \bibinfo{author}{\bibfnamefont{B.}~\bibnamefont{Londo\~no Flor\`ez}},
  \bibinfo{author}{\bibfnamefont{R.}~\bibnamefont{Vexiau}},
  \bibinfo{author}{\bibfnamefont{O.}~\bibnamefont{Dulieu}},
  \bibinfo{author}{\bibfnamefont{N.}~\bibnamefont{Bouloufa-Maafa}},
  \bibnamefont{and}
  \bibinfo{author}{\bibfnamefont{E.}~\bibnamefont{Luc-Koenig}},
  \bibinfo{journal}{Phys. Rev. A} \textbf{\bibinfo{volume}{90}},
  \bibinfo{pages}{033413} (\bibinfo{year}{2014}).

\end{thebibliography}

\end{document}